\documentclass[12pt]{article}
\usepackage{amsthm, bm, graphicx, hyperref, mathrsfs, bbm}
\usepackage{amsfonts}
\usepackage{amsmath}
\usepackage{amssymb}
\usepackage{dsfont}
\usepackage{graphicx}
\usepackage{color}
\usepackage[all, knot]{xy}
\usepackage{tikz}
\usepackage{braket}
\usepackage{epstopdf}
\usepackage[footnotesize]{caption}
\usepackage{amsthm}
\usepackage{enumitem}
\usepackage{mathrsfs}
\usepackage{mathtools}     
\usepackage{subeqnarray}         
\usepackage{cases}               
\usepackage{color}
\usepackage{subfigure}
\usepackage{cite}               
\usepackage{hyperref}            
\usepackage{multirow,makecell}   
\usepackage{textcomp}
\usepackage{wasysym}
\usepackage{url}
\usepackage[margin=3cm]{geometry}
\usepackage{geometry}
\usepackage{amsfonts}
\usepackage{graphicx}
\usepackage{float}
\usepackage{amsmath}
\usepackage{hyperref}
\usepackage{tensor}
\usepackage{subfigure}
\usepackage{cases}
\usepackage{appendix}
\usepackage{multirow}
\usepackage{color}
    \DeclareFontFamily{U}{wncy}{}
    \DeclareFontShape{U}{wncy}{m}{n}{<->wncyr10}{}
    \DeclareSymbolFont{mcy}{U}{wncy}{m}{n}
    \DeclareMathSymbol{\Sh}{\mathord}{mcy}{"58} 
\allowdisplaybreaks[3]

\begin{document}
\begin{titlepage}
\vspace{0.5cm}
\begin{center}
{\Large\bf{Holographic torus correlators in AdS$_3$ gravity coupled to scalar field}}
\lineskip .75em
\vskip 2.5cm
{\large{Song He$^{\clubsuit,\maltese,}$\footnote{hesong@jlu.edu.cn}, Yun-Ze Li$^{\clubsuit,}$\footnote{lyz21@mails.jlu.edu.cn}, Yunda Zhang$^{\clubsuit,}$\footnote{ydzhang21@mails.jlu.edu.cn}}}
\vskip 2.5em
{\normalsize\it $^\clubsuit$Center for Theoretical Physics and College of Physics, Jilin University,\\
 Changchun 130012, People's Republic of China
\\$^{\maltese}$Max Planck Institute for Gravitational Physics (Albert Einstein Institute),\\ 
Am M\"uhlenberg 1, 14476 Golm, Germany}
\end{center}
\begin{abstract}
This paper investigates holographic torus correlators of generic operators at conformal infinity and a finite cutoff within AdS$_3$ gravity coupled with a free scalar field. Using a near-boundary analysis and solving the gravitational boundary value problem, we solve Einstein's equation and calculate mixed correlators for massless and massive coupled scalar fields. The conformal Ward identity on the torus has been reproduced holographically, which can be regarded as a consistency check. Further, recurrence relations for a specific class of higher-point correlators are derived, validating AdS$_3$/CFT$_2$ with non-trivial boundary topology. While the two-point scalar correlator is accurately computed on the thermal AdS$_3$ saddle, the higher-point correlators associated with scalar and stress tensor operators are explored.
\end{abstract}
\end{titlepage}
\newpage
\tableofcontents
\section{Introduction}\label{1}
Understanding the nonperturbative effects in strongly coupled physics has been a longstanding challenge. As a concrete realization of the holographic principle \cite{tHooft:1993dmi, Susskind:1994vu}, the Anti-de Sitter gravity/conformal field theory (AdS/CFT) correspondence \cite{Maldacena:1997re, Gubser:1998bc, Witten:1998qj} provides a potent tool for analytically studying strongly coupled field theories. A crucial application of this correspondence lies in obtaining the correlators of local operators in the boundary CFT through gravitational perturbative calculations performed in the bulk. The stress tensor correlators, which contain information about a system's energy, momentum, and stress distribution, have garnered substantial research attention. The holographic computations of the stress tensor correlators have been done in many remarkable works \cite{Liu:1998ty, Arutyunov:1999nw, DHoker:1999bve, Raju:2012zs, Bagchi:2015wna} when considering a boundary CFT with a trivial topology. Simultaneously, it is imperative to investigate holographic correlators on nontrivial topological manifolds as they offer valuable insights into the holographic principle and the behavior of CFTs in curved spacetimes. 

Exploring correlators in CFTs with nontrivial topologies through the variational principle involves solving Einstein's equations in a general bulk geometry. While solutions near the boundary are well-established \cite{Fefferman:1985ci, Henningson:1998gx, Skenderis:1999nb,deHaro:2000vlm, Fefferman:2007rka}, the absence of the maximal symmetry in pure AdS poses a substantial challenge for addressing the global boundary value problem. This difficulty persists even when dealing with linearized equations, as discussed in previous literature \cite{Graham:1991jqw,anderson2004structure, Anderson:2004yi,anderson2008einstein}. 
Moreover, the research community's understanding of higher-point stress tensor correlators in strongly coupled CFTs with nontrivial topology remains limited. There is a pressing need for explicit results obtained through holographic computations to shed light on these complicated correlation functions. 
In our previous research \cite{He:2023hoj}, we delved into the study of holographic stress tensor correlators within the framework of AdS$_3$/CFT$_2$ when considering a CFT defined on a torus. We expanded upon this approach in \cite{He:2023wcs}, extending our analysis to AdS$_5$/CFT$_4$. We computed the Euclidean thermal two-point correlators for the stress tensor and the $U(1)$ current in the latter case. Recent advancements in this field have also led to investigations of correlation functions within holographic CFTs featuring nontrivial topologies, as reported in \cite{Bhatta:2023qcl, Georgiou:2023xpg}. These studies contribute to the ongoing quest for a deeper understanding of holography in diverse CFT settings.\par

When calculating the holographic stress tensor correlators, one may consider a pure gravitational system in the bulk, where the boundary metric is the only source in the corresponding CFT. Extending the computation of correlators to holographic CFTs that contain multiple sources is a natural progression. A crucial research direction involves investigating Einstein's gravity minimally coupled to the matter field. This study focuses on calculating holographic torus correlators in a minimally coupled Einstein-scalar system within the framework of AdS$_3$/CFT$_2$. Before this, holographic correlators of Einstein-scalar theories have been extensively investigated on manifolds with trivial topology \cite{deHaro:2000vlm, Bianchi:2001de,Bianchi:2001kw,Skenderis:2002wp,Papadimitriou:2004ap,Papadimitriou:2004rz}. By considering multiple sources, we can compute holographic correlators that involve a mixed insertion of both the dual scalar operator and the stress tensor. The contribution of the inserted stress tensor is obtained by perturbatively solving Einstein's equation, leading to the derivation of the holographic torus Ward identities. This result serves as a concrete verification of AdS$_3$/CFT$_2$ with non-trivial topology. The derivation of recurrence relations is presented herein to compute higher-point correlators. To obtain the precise form of holographic correlators, we subsequently calculate the two-point scalar correlator on the thermal AdS$_3$ saddle. The procedure for computing holographic correlators is extended to an Einstein-scalar system featuring a boundary at a finite cutoff. From the perspective of field theory, such a bulk theory is dual to $T^2$-deformed CFT\cite{Hartman:2018tkw, Taylor:2018xcy, Shyam:2018sro, Belin:2020oib}.\par
The remainder of this paper is organized as follows. In section \ref{2}, we review fundamental techniques for computing holographic torus correlators. In section \ref{3}, we calculate the holographic correlators on the conformal boundary. We begin by considering the scenario where the coupled scalar field is massless in subsection \ref{3.1}. The recurrence relations on a general saddle are derived in subsubsection \ref{3.1.1}, and the precise two-point scalar correlator on the thermal AdS$_3$ saddle is computed in subsubsection \ref{3.1.2}. In subsection \ref{3.2}, we investigate the scenario where the coupled scalar field is massive and derive the corresponding recurrence relations. In section \ref{4}, we present a method for calculating holographic torus correlators at a finite cutoff. We analyze the two-point correlators $\langle{T_{ij}O}\rangle$ and $\langle{T_{ij}T_{kl}}\rangle$ on a general saddle in subsection \ref{4.1}. The exact three-point correlators $\langle{T_{ij}OO}\rangle$ on the thermal AdS$_3$ saddle is computed in subsection \ref{4.2}. In section \ref{6} we provide a comprehensive summary and future perspectives.

\section{Holographic setup and torus correlators}\label{2}
Let us briefly review the basics of calculating holographic correlators. Our calculations are within the framework of AdS$_3$/CFT$_2$ correspondence. The three-dimensional bulk $\mathcal{M}$ is a solid torus that encompasses Einstein gravity (with a negative cosmological constant) and a free scalar field\footnote{We follow the convention of \cite{deHaro:2000vlm} for the bulk action.},
\begin{align}
    I_{\text{bulk}}=&\frac{1}{16\pi G}\int_{\mathcal{M}}d^3x\sqrt{\mathcal{G}}\Big(\mathcal{R}-2\Big)-\frac{1}{8\pi G}\int_{\partial\mathcal{M}}d^2x\sqrt{\gamma}K\notag\\
    &+\frac{1}{2}\int_{\mathcal{M}}d^3x\sqrt{\mathcal{G}}\Big((\partial\Phi)^2+m^2\Phi^2\Big)+I_{\text{ct}}.\label{3d action}
\end{align}
Here, we have set the AdS radius $l=1$. The second term is the Gibbons-Hawking term \cite{Gibbons:1976ue}, where $K=\gamma^{ij}K_{ij}$ with the induced metric $\gamma_{ij}$ and the extrinsic curvature $K_{ij}$. Einstein's equations and scalar equation of motion (EOM) are obtained from the variations of (\ref{3d action}) w.r.t. $\mathcal{G}^{\mu\nu}$ and $\Phi$,
\begin{align}
    \mathcal{R}_{\mu\nu}-2\mathcal{G}_{\mu\nu}+8\pi G[\partial_{\mu}\Phi\partial_{\nu}\Phi+m^2\Phi^2\mathcal{G}_{\mu\nu}]=&0,\label{Einstein's equations 0}\\
    \Box_{(\mathcal{G})}\Phi-m^2\Phi=&0. \label{scalar EOM}
\end{align}
The Laplacian $\Box_{(\mathcal{G})}\Phi=\frac{1}{\sqrt{\mathcal{G}}}\partial_{\mu}(\sqrt{\mathcal{G}}\mathcal{G}^{\mu\nu}\partial_{\nu}\Phi)$. 
The on-shell action diverges when evaluated at the boundary, requiring the inclusion of the counterterm $I_{\text{ct}}$ in (\ref{3d action}) to eliminate the divergence. This prescription is called holographic renormalization \cite{deHaro:2000vlm} (see also \cite{Henningson:1998gx,Bianchi:2001kw,Skenderis:2002wp,Balasubramanian:1999re,Emparan:1999pm,Kraus:1999di}) and we will review it in appendix \ref{2.1}.\par
It is convenient to work in the Fefferman-Graham (FG) coordinate system \cite{fefferman1985conformal,2007arXiv0710.0919F}, in which the bulk metric $\mathcal{G}_{\mu\nu}$ takes the form 
\begin{align}
    ds^2=\mathcal{G}_{\mu\nu}dx^{\mu}dx^{\nu}=\frac{d\rho^2}{4\rho^2}+\frac{1}{\rho}g_{ij}(\rho,x)dx^idx^j.\label{Fefferman-Graham coordinates}
\end{align}
The radial coordinate $\rho\in[0,\pi^{-2}]$ and $\rho=0$ is the conformal boundary. One can introduce the complex tangential coordinates\footnote{Throughout this paper, we work with Euclidean signature.} $(x^1,x^2)=(z,\bar z)$ with identifications $(z,\bar z)\cong (z+1,\bar z+1)\cong (z+\tau,\bar z+\bar\tau)$, where $\tau$ is the modular parameter of the conformal torus boundary. In the FG coordinate system, bulk solutions $g_{ij}(\rho,x)$ and $\Phi(\rho,x)$ can be constructed near the conformal boundary. As discussed in \cite{Hung:2011ta} (see also \cite{Witten:1998qj,deHaro:2000vlm}), there are two homogeneous solutions $g^-_{ij}$ and $\rho g^+_{ij}$ for the metric, and two homogeneous solutions $\rho^{1-\frac{\Delta}{2}}\phi^{-}$ and $\rho^{\frac{\Delta}{2}}\phi^+$ for the scalar field, where $\Delta=1+\sqrt{1+m^2}$ is the scaling dimension of the dual operator. The bulk solutions can be formally written as
\begin{align}
    g_{ij}(\rho,x)=&g^{-}_{ij}(\rho,x)+\rho g^{+}_{ij}(\rho,x),\notag\\
    \phi(\rho,x)=&\phi^{-}(\rho,x)+\rho^{\Delta-1}\phi^+(\rho,x),\label{two homogeneous solutions}
\end{align}
where $\phi(\rho,x)=\rho^{\frac{\Delta}{2}-1}\Phi(\rho,x)$. The leading order of each solution has a specific interpretation. $g^{-}_{ij}(0,x)$ and $\phi^{-}(0,x)$ are the sources couple to the boundary stress tensor $T^{ij}$ and the dual scalar operator $O$ respectively. $g^{+}_{ij}(0,x)$ and $\phi^{+}(0,x)$ are related to one-point correlators $\langle{T_{ij}}\rangle$ and $\langle{O}\rangle$ in the boundary CFT.\par
By appropriately choosing the scaling dimension, each solution in (\ref{two homogeneous solutions}) can be written as a series expansion of $\rho$, called the FG expansion. The cases for $\Delta=1,2,\cdots$ have been considered in \cite{deHaro:2000vlm}, where the series expansion contains the integer powers of $\rho$ and the logarithmic terms. In particular, $\Delta=2$ corresponds to massless scalar field coupling whose FG expansions take the forms
\begin{align}
    g_{ij}(\rho,x)=&\sum_{n=0}^\infty\rho^{n}\Big[g_{(2n)ij}+\rho\text{ln}\rho h_{(2n+2)ij}\Big](x),\notag\\
    \phi(\rho,x)=&\sum_{n=0}^\infty\rho^{n}\Big[\phi_{(2n)}+\rho\text{ln}\rho \psi_{(2n+2)}\Big](x). \label{massless FG expansion}
\end{align}
The expansions of two homogeneous solutions $g^{-}_{ij}$ and $\rho g^{+}_{ij}$ overlap on the positive integer orders of $\rho$, which requires the presence of the logarithmic terms $\rho^n\text{ln}\rho h_{(2n)ij}(x)$ for $n\geq 1$. Similarly, the logarithmic terms $\rho^n\text{ln}\rho\psi_{(2n)}(x)$ for $n\geq 1$ are added to the scalar field expansion. The zeroth-order coefficients $g_{(0)ij}(x)$ and $\phi_{(0)}(x)$ (corresponding to $g^{-}_{ij}(0,x)$ and $\phi^{-}(0,x)$ above) are the boundary values of the metric and scalar field. The first-order coefficients $g_{(2)ij}(x)$ and $\phi_{(2)}(x)$ (corresponding to $g^{+}_{ij}(0,x)$ and $\phi^{+}(0,x)$ above) are determined by the state of boundary CFT, which are independent of the boundary values. Plugging (\ref{Fefferman-Graham coordinates})(\ref{massless FG expansion}) into (\ref{Einstein's equations 0})(\ref{scalar EOM}) and solving the EOMs order by order, one obtains each coefficient of (\ref{massless FG expansion}) in terms of $g_{(0)ij}(x)$, $\phi_{(0)}(x)$, $g_{(2)ij}(x)$ and $\phi_{(2)}(x)$. Especially, $h_{(2)ij}(x)$ and $\psi_{(2)}(x)$ correspond to the conformal anomalies of boundary CFT, which contribute to one-point correlators $\langle{T_{ij}}\rangle$ and $\langle{O}\rangle$ respectively \cite{Papadimitriou:2004rz,deHaro:2000vlm, Petkou:1999fv}. More generally, for the case where $\Delta$ is a rational number, the fractional powers of $\rho$ appear in the FG expansions \cite{Hung:2011ta}, which will be discussed in section \ref{3.2}.\par
In the case of massless scalar field coupling, one-point correlators are obtained through holographic renormalization \cite{deHaro:2000vlm},
\begin{align}
    \langle{T_{ij}}\rangle=&\frac{1}{8\pi G}\Big(g_{(2)ij}-4\pi G\partial_i\phi_{(0)}\partial_j\phi_{(0)}-2\pi G \partial^k\phi_{(0)}\partial_k\phi_{(0)}g_{(0)ij}-\frac{1}{2}R_{(0)}g_{(0)ij}\Big),\notag\\
    \langle{O}\rangle=&2\Big(\phi_{(2)}-\frac{1}{4}\Box_{(0)}\phi_{(0)}\Big), \label{massless case one-point correlators}
\end{align}
where the indices $i,j,k,l$ are raised by $g^{ij}_{(0)}$. $\langle{T_{ij}}\rangle$ has the following trace relation and conservation equation,
\begin{align}
    \langle{T^i_{i}}\rangle=&-\frac{1}{16\pi G}R_{(0)}-\frac{1}{2}\partial^i\phi_{(0)}\partial_i\phi_{(0)},\label{massless trace relation}\\
    \nabla_{(0)}^j\langle{T_{ij}}\rangle=&\langle{O}\rangle\partial_i\phi_{(0)}.\label{massless conservation equation}
\end{align}
Detailed derivation is reviewed in appendix \ref{2.1}.
\subsection{Prescription of holographic torus correlators}\label{2.2}
Throughout this paper, we primarily calculate holographic correlators within the framework of AdS$_3$/CFT$_2$ correspondence. This correspondence is characterized by the Gubser-Klebanov-Polyakov-Witten (GKPW) relation \cite{Gubser:1998bc, Witten:1998qj}, which establishes the equivalence between the bulk gravitational partition function and the boundary generating functional,
\begin{align}
    Z_{G}[\phi_{(0)},g_{(0)ij}]=\Big\langle{\text{exp}\int_{\partial\mathcal{M}}d^2x\sqrt{g_{(0)}}(\phi_{(0)}O-\frac{1}{2}g^{ij}_{(0)}T_{ij})}\Big\rangle_{\text{CFT}}.\label{GKPW rekation}
\end{align}
In the semiclassical limit, the gravitational partition function can be approximated as a sum over all saddles, $Z_{G}[\phi_{(0)},g_{(0)ij}]\approx\sum_{\alpha}e^{-I^{(\alpha)}_{\text{on-shell}}[\phi_{(0)},g_{(0)ij}]}$. Assuming that the partition function $Z_{G}[\phi_{(0)},g_{(0)ij}]$ is dominated by only one saddle, holographic correlators can be obtained by taking functional derivatives of its on-shell action \cite{Skenderis:2002wp, Papadimitriou:2004ap, Kraus:2018xrn, Li:2020pwa},
\begin{align}
    &\Big\langle{\prod_{k=1}^m T_{i_kj_k}(x_k)\prod_{l=1}^n O(y_l)}\Big\rangle_{c}\notag\\
    =&-\Big(\prod_{k=1}^m\frac{-2}{\sqrt{g_{(0)}(x_k)}}\Big)\Big(\prod_{l=1}^n\frac{1}{\sqrt{g_{(0)}(y_l)}}\Big)\Big(\prod_{k=1}^m\frac{\delta}{\delta g^{i_kj_k}_{(0)}(x_k)}\prod_{l=1}^n\frac{\delta}{\delta\phi_{(0)}(y_l)}\Big)I_{\text{on-shell}}\Big|_{\substack{\phi_{(0)}=0\\g_{(0)ij}=\eta_{ij}}} ,\label{holographic correlator}
\end{align}
where the subscript $c$ on the left-hand side implies the connected part of the correlator. The functional derivative is evaluated at the fixed point $(\phi_{(0)},g_{(0)ij})=(0,\eta_{ij})$. 
We are considering three types of holographic correlators. For the case where $m\neq 0$ and $n=0$, we obtain $m$-point stress tensor correlators, which have been computed in a prior study \cite{He:2023hoj}. When $m=0$ and $n\neq 0$, we obtain $n$-point scalar correlators, as discussed in Section \ref{3.1.2}. However, our primary focus lies on the case where $m\neq 0$ and $n\neq 0$, resulting in $(m+n)$-point mixed correlators. Throughout this paper, we will calculate mixed correlators in various scenarios and compare them with results in torus CFT.\par
To calculate holographic correlators, we introduce perturbations $\delta\phi_{(0)}$ and $\delta g_{(0)ij}$ to the boundary values and then solve the bulk equations of motion (\ref{Einstein's equations 0}) and (\ref{scalar EOM}) with the perturbed Dirichlet boundary conditions. However, the bulk fields are constructed by both the boundary values ($\phi_{(0)}$ and $g_{(0)ij}$) and the state-dependent data ($\phi_{(2)}$ and $g_{(2)ij}$). When perturbing the boundary values, the changes $\delta\phi_{(2)}$ and $\delta g_{(2)ij}$ in the state-dependent data are not uniquely determined. In previous work \cite{He:2023hoj}, we employed the global regularity condition to constrain the remaining degrees of freedom. The crucial requirement is that the perturbed bulk solutions described by the FG expansions near the boundary can be smoothly extended to global solutions. As mentioned earlier, the bulk $\mathcal{M}$ is a solid torus, and the cutoff surface at $\rho=\rho_c$ is a torus with the induced metric $\gamma_{ij}(\rho_c,x)$. As $\rho_c$ approaches $\pi^{-2}$, the cutoff torus shrinks into a circle and the bulk fields must be regular here. For the perturbed scalar field, we have
\begin{align}
    \lim_{\rho\to \pi^{-2}}\delta\Phi(\rho,x)=\delta\Phi^*(x)\ \ \ \text{finite},\label{scalar global regularity condition}
\end{align}
which plays a crucial role in section \ref{3.1.2}. The global regularity condition for the perturbed bulk metric $\delta g_{ij}(\rho,x)$ is more complicated. After changing the bulk metric, the new FG coordinates differ from the original FG coordinates by a boundary preserving diffeomorphism\footnote{For a specific form of this boundary preserving diffeomorphism, please refer to Appendix B of \cite{He:2023hoj}.}. Then, the perturbed bulk metric can be divided into two parts: the change by the boundary preserving diffeomorphism and the variation in the new FG coordinate system. Transforming them into a smoothly compatible chart covering the circle at $\rho=\pi^{-2}$ and requiring all its components to be globally regular, one can obtain some constraints on $\delta g_{ij}(\pi^{-2},x)$. In previous work~\cite{He:2023hoj}, we have used these constraints to fix the undetermined constants in two-point stress tensor correlators. For higher-point correlators, the calculation of the constraints becomes increasingly tedious.\par
There is an alternative method to fix the remaining degrees of freedom in holographic correlators \cite{Friedan:1986ua, Eguchi:1986sb} (see also section 7.2 of \cite{Polchinski:1998rq}). For the correlator with at least one stress tensor insertion, the undetermined part can always be written as a one-point-averaged correlator\footnote{In this paper, the complex variable $z$ collectively represents both $z$ and its complex conjugate $\bar z$.},
\begin{align}
    \int_{\partial\mathcal{M}}d^2z\langle{T^{ij}(z)X}\rangle=\int_{\partial\mathcal{M}}d^2z\Big(\frac{\delta\langle{X}\rangle}{\delta g_{(0)ij}(z)}+\text{contact terms}\Big). \label{one-point-averaged correlator}
\end{align}
Let us calculate the first term on the right-hand side. Consider the global variations $\delta g_{(0)\bar z\bar z}=\alpha$ and $\delta g_{(0)zz}=\bar\alpha$ on the boundary metric, we have
\begin{align}
    ds^2=&dzd\bar z+\alpha d\bar z^2+\bar\alpha dz^2\notag\\
    =&(1+\alpha+\bar\alpha)d[z+\alpha(\bar z-z)]d[\bar z+\bar\alpha(z-\bar z)]+O(\alpha^2).
\end{align}
After performing the global Weyl transformation $g_{(0)ij}\mapsto(1-\alpha-\bar\alpha)g_{(0)ij}$ plus the coordinate transformation $(z',\bar z')=(z+\alpha(\bar z-z),\bar z+\bar\alpha(z-\bar z))$, the torus metric (at leading order) changes back to the Euclidean one $ds^2=dz'd\bar z'+O(\alpha^2)$ with the new modular parameter $\tau'=\tau+\alpha(\bar\tau-\tau)$. For a general correlator, the changes under the global metric variation and the global Weyl transformation are
\begin{align}
(\delta\langle{X}\rangle)_{\text{metric}}=&\alpha\int_{\partial\mathcal{M}}d^2z\frac{\delta\langle{X}\rangle}{\delta g_{(0)\bar z\bar z}(z)}+\bar\alpha\int_{\partial\mathcal{M}}d^2z\frac{\delta\langle{X}\rangle}{\delta g_{(0)zz}(z)},\notag\\
    (\delta\langle{X}\rangle)_{\text{Weyl}}=&-(\alpha+\bar\alpha)\int_{\partial\mathcal{M}}d^2z\frac{\delta\langle{X}\rangle}{\delta g_{(0)z\bar z}(z)}.
\end{align}
The change of $\langle{X}\rangle$ under the coordinate transformation $(z,\bar z)\mapsto (z',\bar z')$ is
\begin{align}
    (\delta\langle{X}\rangle)_{\text{coordinate}}=&\alpha\mathcal{L}_{(z-\bar z)\partial_z}\langle{X}\rangle+\bar\alpha\mathcal{L}_{(\bar z-z)\partial_{\bar z}}\langle{X}\rangle,
\end{align}
where $\mathcal{L}$ denotes the Lie derivative. The sum of the above three transformations is equivalent to varying the modular parameter $\delta\tau=\alpha(\bar\tau-\tau)$, which gives the following two relations~\cite{He:2023hoj},
\begin{align}
    (\bar{\tau}-\tau) \partial_\tau \langle{X}\rangle &=\mathcal{L}_{(z-\bar{z})\partial_z}\langle{X}\rangle
    +\int_{\partial\mathcal{M}} d^2z \Big( \frac{\delta\langle{X}\rangle}{\delta g_{(0)\bar{z}\bar{z}}(z)} -  \frac{\delta\langle{X}\rangle}{\delta g_{(0)z\bar{z}}(z)} \Big),\notag\\
    (\tau-\bar{\tau}) \partial_{\bar{\tau}} \langle{X}\rangle &= \mathcal{L}_{(\bar{z}-z)\partial_{\bar{z}}}\langle{X}\rangle
    + \int_{\partial\mathcal{M}} d^2z \Big( \frac{\delta\langle{X}\rangle}{\delta g_{(0)zz}(z)} -  \frac{\delta\langle{X}\rangle}{\delta g_{(0)z\bar{z}}(z)}\Big). \label{alternative method}
\end{align}
Later, we will utilize these relations to compute the two-point and three-point mixed correlators and derive recurrence relations for higher-point mixed correlators.
\section{Correlators in holographic CFT}\label{3}
In this section, holographic correlators on the conformal boundary are calculated in two cases: the coupled scalar field is massless ($\Delta=2$) and massive ($1\leq \Delta <2$). We mainly focus on obtaining mixed correlators. The basic idea is to perturb the boundary values $\phi_{(0)}$ and $g_{(0)ij}$ and solve the trace relation and the conservation equation order by order. Furthermore, we compute the two-point scalar correlator and give a preliminary discussion of the higher-point scalar correlators.
\subsection{Massless scalar field}\label{3.1}
\subsubsection{Perturbative solution of massless scalar EOM}\label{3.1.2}
As we perturb the boundary value of the scalar field $\Phi_{(0)}(x)=\epsilon f(x)$, the bulk metric $\mathcal{G}^{[\epsilon]}_{\mu\nu}$ also varies due to the back-reaction. Hence, the scalar EOM should be considered in a changing background,
\begin{align}
    \frac{1}{\sqrt{\mathcal{G}^{[\epsilon]}}}\partial_{\mu}\Big(\sqrt{\mathcal{G}^{[\epsilon]}}\mathcal{G}^{[\epsilon]\mu\nu}\partial_{\nu}\Phi^{[\epsilon]}\Big)=0. \label{massless scalar EOM with changing background}
\end{align}
The perturbed bulk fields have Taylor expansions $\Phi^{[\epsilon]}=\sum_{k=0}^{\infty}\epsilon^k\Phi^{[k]}$ and $\mathcal{G}_{\mu\nu}^{[\epsilon]}=\sum_{k=0}^{\infty}\epsilon^k\mathcal{G}_{\mu\nu}^{[k]}$. Plugging them into (\ref{massless scalar EOM with changing background}), we read off the differential equation for $\Phi^{[k]}$. Meanwhile, $\Phi^{[k]}$ has the FG expansion in the coordinate system (\ref{Fefferman-Graham coordinates}), and the subleading term $\Phi^{[k]}_{(2)}$ corresponds to the one-point scalar correlator $\langle{O}\rangle^{[k]}$. The $k$-th order variation of $\langle{O}\rangle^{[k]}$ gives the $(n+1)$-point scalar correlator in the boundary CFT.\par
The zeroth-order coefficients $\Phi^{[0]}$ and $\mathcal{G}_{\mu\nu}^{[0]}$ are given by the saddle before perturbation. In the FG coordinate system (\ref{Fefferman-Graham coordinates}), they satisfy the boundary conditions\footnote{In fact, the boundary condition of the massless scalar field can be extended to $\Phi^{[0]}|_{\rho\to 0}=C$, where $C$ is a constant. The boundary field theory can be regarded as a marginal deformation of the original CFT,
\begin{align}
S'=S_{CFT}+C\int_{\partial\mathcal{M}} d^2zO.
\end{align}
This class of boundary conditions will lead to the same holographic Ward identities.} $\Phi^{[0]}|_{\rho\to 0}=0$ and $\rho\mathcal{G}_{ij}^{[0]}|_{\rho\to 0}=\eta_{ij}$. The simplest type of saddle is characterized by the disappearance of the bulk scalar field. For instance, the thermal AdS$_3$ solution
\begin{align}
    ds^2=\mathcal{G}^{[0]}_{\mu\nu}dx^\mu dx^\nu=\frac{1}{1+r^2}dr^2+r^2d\varphi^2+(1+r^2)dt^2,\ \ \ \Phi^{[0]}(r,\varphi,t)=0,\label{metric thermal AdS3}
\end{align}
where $(\varphi,t)$ are the real tangential coordinates with $\varphi\sim\varphi+2\pi$ and $t\sim t+2\pi T$. The radial coordinate $r\in[0,\infty)$ and $r\to\infty$ is the conformal infinity. Such kind of pure gravitational saddle is considered trivial since it does not contribute to the one-point scalar correlator of the dual CFT.\par
We further compute the two-point scalar correlator of the saddle (\ref{metric thermal AdS3}). 
The perturbed scalar field $\Phi^{[\epsilon]}$ satisfies the Dirichlet boundary condition at conformal infinity and regularity condition at $r=0$,
\begin{align}
    \Phi^{[\epsilon]}(r,\varphi,t)\Big|_{r\to\infty}&=\epsilon f(\varphi,t),\label{conformal boundary}\\
    \Phi^{[\epsilon]}(r,\varphi,t)\Big|_{r\to 0}&\ \ \ \ \text{regular}.\label{global regularity condition}
\end{align}
Since $\Phi^{[0]}$ is turned off, the first-order EOM is a homogeneous differential equation,
\begin{align}
    \frac{1}{\sqrt{\mathcal{G}^{[0]}}}\partial_{\mu}\Big(\sqrt{\mathcal{G}^{[0]}}\mathcal{G}^{[0]\mu\nu}\partial_{\nu}\Phi^{[1]}\Big)=0.\label{1-st scalar EOM}
\end{align}
The investigation of scalar field solutions within a fixed gravitational background has been extensively explored in numerous works \cite{Ichinose:1994rg,Freedman:1998tz,Cardoso:2001hn,Birmingham:2001pj,Birmingham:2001hc,Lopez-Ortega:2018efa,Dodelson:2022yvn,Lei:2023mqx,Bhatta:2023qcl,Bhatta:2022wga,Harlow:2011ke}. The majority of these studies utilize the Minkowski signature and implement the ingoing boundary condition at the horizon. The derivation in the Euclidean signature space is analogous. The double periodicity of $\Phi^{[1]}$ allows us to represent it formally as a Fourier expansion,
\begin{align}
    \Phi^{[1]}(r,\varphi,t)=&\sum_{m,n=-\infty}^{+\infty}\Phi^{[1]}_{m,n}(r)e^{\frac{-imt}{T}}e^{in\varphi}.\label{Fourier expansion}
\end{align}
Plugging (\ref{metric thermal AdS3})(\ref{Fourier expansion}) into (\ref{1-st scalar EOM}), we obtain
\begin{align}
    \frac{1}{r}\frac{d}{dr}\Big(r(1+r^2)\frac{d}{dr}\Phi^{[1]}_{m,n}\Big)+\Big(\frac{-m^2}{(1+r^2)T^2}-\frac{n^2}{r^2}\Big)\Phi^{[1]}_{m,n}=0.\label{1-st scalar radial equation}
\end{align}
Applying the following coordinate transformation and field redefinition,
\begin{align}
x=1-\frac{1}{1+r^2},\ \ \ \ \Phi^{[1]}_{m,n}(r)\to x^{\frac{n}{2}}(1-x)y^{[1]}_{m,n}(x),\label{transformation}
\end{align}
we rewrite (\ref{1-st scalar radial equation}) in the form of hypergeometric equation,
\begin{align}
    x(1-x)\frac{d^2}{dx^2}y^{[1]}_{m,n}+[\gamma-(\alpha+\beta+1)x]\frac{d}{dx}y^{[1]}_{m,n}-\alpha\beta y^{[1]}_{m,n}=0,\label{hypergeometric equation}
\end{align}
where the parameters $\alpha=1+\frac{im}{2T}+\frac{n}{2}$, $\beta=1-\frac{im}{2T}+\frac{n}{2}$, and $\gamma=1+n$. The new radial coordinate $x\in [0,1]$ with the conformal boundary at $x=1$ and the circle at $x=0$. The boundary conditions for $y^{[1]}_{m,n}$ are read off from (\ref{conformal boundary})(\ref{global regularity condition}),
\begin{align}
(1-x)y^{[1]}_{m,n}(x)\Big|_{x\to 1}&=f_{m,n},\label{y's BC}\\
    x^{\frac{n}{2}}y^{[1]}_{m,n}(x)\Big|_{x\to 0}&\ \ \ \ \text{regular},\label{y's GRC}
\end{align}
where $f_{m,n}=\frac{1}{4\pi^2T}\int_{-\pi}^{\pi}d\varphi\int_{-\pi T}^{\pi T}dtf(\varphi,t)e^{\frac{imt}{T}}e^{-in\varphi}$ is the Fourier coefficient of $f(\varphi,t)$. Equation (\ref{hypergeometric equation}) has two independent solutions in the neighborhood of $x=0$, and only one of them satisfies the regularity condition (\ref{y's GRC}). The form of the allowed solution varies depending on the sign of $n$,
 \begin{gather}
y^{[1]}_{m,n}(x)=
\begin{cases}
A^{[1]}_{m,n}F(\alpha,\beta,\gamma,x),&n\geq 0,\\
A^{[1]}_{m,n}x^{-n}F(\alpha-n,\beta-n,1-n,x),&n<0.
\end{cases} \label{allowed solution}
\end{gather}
$F(\alpha,\beta,\gamma,x)$ is the hypergeometric function \cite{abramowitz1968handbook} (and note that $\gamma-\alpha-\beta=-1$), which has the following expansion in the neighborhood of $x=1$,
\begin{align}
    F(\alpha,\beta,\gamma,x)=&\frac{\Gamma(\gamma)(1-x)^{-1}}{\Gamma(\alpha)\Gamma(\beta)}+\frac{\Gamma(\gamma)}{\Gamma(\alpha-1)\Gamma(\beta-1)}\Big[\text{ln}(1-x)\notag\\
    &+[\psi(\alpha)+\psi(\beta)+2\gamma_E-1]\Big]+O(1-x), \label{F's expansion at x=1}
\end{align}
where $\psi(\lambda)=\frac{\Gamma'(\lambda)}{\Gamma(\lambda)}$ is the logarithmic differentiation of the Gamma function and $\gamma_E$ is the Euler constant. Plugging (\ref{allowed solution})(\ref{F's expansion at x=1}) into (\ref{y's BC}), we find
\begin{gather}
A^{[1]}_{m,n}=
\begin{cases}
\frac{\Gamma(\alpha)\Gamma(\beta)}{\Gamma(\gamma)}f_{m,n},&n\geq 0,\\
\frac{\Gamma(\alpha-n)\Gamma(\beta-n)}{\Gamma(1-n)}f_{m,n},&n<0.
\end{cases} \label{constant A1mn}
\end{gather}
Putting everything together, we obtain the global solution $\Phi^{[1]}$, then we expand it around the conformal boundary,
\begin{align}
    &\Phi^{[1]}(x,\varphi,t)\notag\\
    =&\sum_{m,n=-\infty}^{+\infty}\Big\lbrace1+\Big[(\frac{m^2}{4T^2}+\frac{n^2}{4})[\psi(1+\frac{im}{2T}+\frac{|n|}{2})+\psi(1-\frac{im}{2T}+\frac{|n|}{2})+2\gamma_E-1]-\frac{|n|}{2}\Big](1-x)\notag\\
    &+(\frac{m^2}{4T^2}+\frac{n^2}{4})(1-x)\text{ln}(1-x)\Big\rbrace f_{m,n}e^{\frac{-imt}{T}}e^{in\varphi}+O((1-x)^2). \label{Phi 1 real solution}
\end{align}
Now, let us return to the FG coordinate system (\ref{Fefferman-Graham coordinates}). Applying the following coordinate transformation
\begin{align}
    x=\Big(\frac{1-\pi^2\rho}{1+\pi^2\rho}\Big)^2,\ \ \ \ \varphi=\pi(z+\bar z),\ \ \ \ t=-i\pi(z-\bar z), \label{real coordinates to FG patch}
\end{align}
the bulk metric of the thermal AdS$_3$ can be rewritten as
\begin{align}
    ds^2=&\frac{d\rho^2}{4\rho^2}+\frac{1}{\rho}\Big[dzd\bar z-\pi^2\rho(dz^2+d\bar z^2)+\pi^4\rho^2dzd\bar z\Big], \label{bulk metric of the thermal AdS}
\end{align}
and the FG coefficients of $\Phi^{[1]}$ take the forms
\begin{align}
    \Phi^{[1]}_{(0)}(z)=&f(z),\notag\\
    \Phi^{[1]}_{(2)}(z)=&4\pi^2\sum_{m,n=-\infty}^{+\infty}\Big[(\frac{m^2}{4T^2}+\frac{n^2}{4})[\psi(1+\frac{im}{2T}+\frac{|n|}{2})+\psi(1-\frac{im}{2T}+\frac{|n|}{2})\notag\\
    &+2\gamma_E-1+\text{ln}(4\pi^2)]-\frac{|n|}{2}\Big]f_{m,n}e^{i\pi(n+\frac{im}{T})z}e^{i\pi(n-\frac{im}{T})\bar z},\notag\\
    \Psi^{[1]}_{(2)}(z)=&4\pi^2\sum_{m,n=-\infty}^{+\infty}(\frac{m^2}{4T^2}+\frac{n^2}{4})f_{m,n}e^{i\pi(n+\frac{im}{T})z}e^{i\pi(n-\frac{im}{T})\bar z}.\label{real FG coefficients}
\end{align}
By definition, the perturbed one-point correlator $\langle{O}\rangle^{[1]}=2(\Phi^{[1]}_{(2)}+\Psi^{[1]}_{(2)})$. Taking the functional derivative w.r.t. $f(z')$, we obtain the two-point correlator,
\begin{align}
    \langle{O(z)O(z')}\rangle=&\frac{8\pi^2}{T}\sum_{m,n=-\infty}^{+\infty}\Big[(\frac{m^2}{4T^2}+\frac{n^2}{4})[\psi(1+\frac{im}{2T}+\frac{|n|}{2})+\psi(1-\frac{im}{2T}+\frac{|n|}{2})\notag\\
    &+2\gamma_E+\text{ln}(4\pi^2)]-\frac{|n|}{2}\Big]e^{i\pi(n+\frac{im}{T})(z-z')}e^{i\pi(n-\frac{im}{T})(\bar z-\bar z')}. \label{scalar two-point correlator}
\end{align}
An alternative method for obtaining the scalar two-point correlator is presented herein. We start from the Poincare AdS$_3$ with the metric $ds^2=\frac{dy^2}{4y^2}+\frac{dwd\bar w}{y}$. Applying the following coordinate transformation
\begin{align}
y=\frac{4\pi^2\rho}{(1+\pi^2\rho)^2}e^{-i2\pi (z-\bar z)},\ \ \ w=\frac{1-\pi^2\rho}{1+\pi^2\rho}e^{-i2\pi z},\ \ \ \bar w=\frac{1-\pi^2\rho}{1+\pi^2\rho}e^{i2\pi \bar z},
\end{align}
we recover the bulk metric (\ref{bulk metric of the thermal AdS}). The calculation of the scalar two-point correlator involves perturbing the boundary value of the scalar field and subsequently solving for the subleading order $\Phi^{[1]}(\rho,z,\bar z)$ from (\ref{1-st scalar EOM}). Consider the following Fourier mode of the scalar field,
\begin{align}
    \Phi^{[1]}_{\omega,k}(x)e^{-i\omega t}e^{ik\varphi}, \label{ansatz for the scalar field}
\end{align}
where the coordinates $(x,t,\varphi)$ are defined in (\ref{real coordinates to FG patch}), and $(\omega,k)$ are real parameters. Plugging (\ref{ansatz for the scalar field}) into (\ref{1-st scalar EOM}) and using the boundary conditions
\begin{align}
    &\Phi^{[1]}_{\omega,k}(x)\Big|_{x\to 1}=f_{\omega,k},\notag\\
    &\Phi^{[1]}_{\omega,k}(x)\Big|_{x\to 0}\ \ \ \ \text{regular},
\end{align}
we obtain
\begin{align}
    \Phi^{[1]}_{\omega,k}(x)=&\frac{\Gamma(1+\frac{i\omega}{2}+\frac{|k|}{2})\Gamma(1-\frac{i\omega}{2}+\frac{|k|}{2})}{\Gamma(1+|k|)}f_{\omega,k}x^{\frac{|k|}{2}}(1-x)\notag\\
   &\times F(1+\frac{i\omega}{2}+\frac{|k|}{2},1-\frac{i\omega}{2}+\frac{|k|}{2},1+|k|,x).
\end{align}
Integrating (\ref{ansatz for the scalar field}) over the momentum space, we find
\begin{align}
    \Phi^{[1]}(x,t,\varphi)=&\int_{\mathbbm{R}^2} d\omega dk\Phi^{[1]}_{\omega,k}(x)e^{-i\omega t}e^{ik\varphi}\notag\\
    =&\int_{\mathbbm{R}^2} dt'd\varphi'K_{\text{g}}(x,t-t',\varphi-\varphi')f(t',\varphi'),
\end{align}
where the bulk-boundary propagator takes the form
\begin{align}
    K_{\text{g}}(x,t-t',\varphi-\varphi')=&\frac{1}{4\pi^2}\int_{\mathbbm{R}^2}d\omega dk\Big[\frac{\Gamma(1+\frac{i\omega}{2}+\frac{|k|}{2})\Gamma(1-\frac{i\omega}{2}+\frac{|k|}{2})}{\Gamma(1+|k|)}x^{\frac{|k|}{2}}(1-x)\notag\\
    &\times F(1+\frac{i\omega}{2}+\frac{|k|}{2},1-\frac{i\omega}{2}+\frac{|k|}{2},1+|k|,x)e^{-i\omega(t-t')}e^{ik(\varphi-\varphi')}\Big]. \label{global AdS propagator}
\end{align}
Now let us return to the thermal AdS$_3$ spacetime, which can be regarded as a quotient of the global AdS$_3$ spacetime \cite{Carlip:1994gc,Kraus:2006wn}. The thermal AdS$_3$ propagator, denoted as $K_{\text{t}}$, satisfies the following periodicity conditions:
\begin{align}
    K_{\text{t}}(x,t-t',\varphi-\varphi')=K_{\text{t}}(x,t-t'+2\pi T,\varphi-\varphi')=K_{\text{t}}(x,t-t',\varphi-\varphi'+2\pi).
\end{align}
This propagator can be constructed from the global AdS$_3$ propagator $K_{\text{g}}$ via the method of images \cite{Shiraishi:1993qnr,Lifschytz:1993eb,Ichinose:1994rg,Keski-Vakkuri:1998gmz}. We shift the Euclidean time $t$ and the angle coordinate $\varphi$ in (\ref{global AdS propagator}) by integer multiples of $2\pi T$ and $2\pi$, respectively, followed by summation over all such propagators to obtain
\begin{align}
    K_{\text{t}}(x,t-t',\varphi-\varphi')=&\sum_{m,n=-\infty}^{\infty}K_{\text{g}}(x,t-t'+2\pi mT,\varphi-\varphi'+2\pi n)\notag\\
    =&\frac{1}{4\pi^2}\int_{\mathbbm{R}^2}d\omega dk\Big[\frac{\Gamma(1+\frac{i\omega}{2}+\frac{|k|}{2})\Gamma(1-\frac{i\omega}{2}+\frac{|k|}{2})}{\Gamma(1+|k|)}x^{\frac{|k|}{2}}(1-x)\notag\\
    &\times F(1+\frac{i\omega}{2}+\frac{|k|}{2},1-\frac{i\omega}{2}+\frac{|k|}{2},1+|k|,x)e^{-i\omega(t-t')}e^{ik(\varphi-\varphi')}\notag\\
    &\times\sum_{m,n=-\infty}^{\infty}e^{-i2\pi mT\omega}e^{i2\pi nk}\Big]. \label{propagator in thermal AdS3}
\end{align}
Consider the Dirac comb function defined as $\Sh_{L}(u)=\sum_{n=-\infty}^{\infty}\delta(u-nL)$, which has the Fourier series expansion
\begin{align}
    \Sh_{L}(u)=\frac{1}{L}\sum_{n=-\infty}^{\infty}e^{i2\pi n\frac{u}{L}}.\label{Dirac comb function}
\end{align}
Plugging (\ref{Dirac comb function}) into (\ref{propagator in thermal AdS3}) we obtain
\begin{align}
    K_{\text{t}}(x,t-t',\varphi-\varphi')=&\frac{1}{4\pi^2T}\sum_{m,n=-\infty}^{\infty}\Big[\frac{\Gamma(1+\frac{im}{2T}+\frac{|n|}{2})\Gamma(1-\frac{im}{2T}+\frac{|n|}{2})}{\Gamma(1+|n|)}x^{\frac{|n|}{2}}(1-x)\notag\\
    &\times F(1+\frac{im}{2T}+\frac{|n|}{2},1-\frac{im}{2T}+\frac{|n|}{2},1+|n|,x)e^{-i\frac{m}{T}(t-t')}e^{in(\varphi-\varphi')}\Big].
\end{align}
It follows that
\begin{align}
    \Phi^{[1]}(x,t,\varphi)=&\int_{-\pi T}^{\pi T}dt'\int_{-\pi}^{\pi}d\varphi' K_{\text{t}}(x,t-t',\varphi-\varphi')f(t',\varphi')\notag\\
    =&\sum_{m,n=-\infty}^{\infty}\Big[\frac{\Gamma(1+\frac{im}{2T}+\frac{|n|}{2})\Gamma(1-\frac{im}{2T}+\frac{|n|}{2})}{\Gamma(1+|n|)}x^{\frac{|n|}{2}}(1-x)\notag\\
    &\times F(1+\frac{im}{2T}+\frac{|n|}{2},1-\frac{im}{2T}+\frac{|n|}{2},1+|n|,x)e^{-i\frac{m}{T}t}e^{in\varphi}f_{m,n}\Big].
\end{align}
This solution is consistent with the one derived from the previous approach. Once again, we expand $\Phi^{[1]}$ around the boundary and extract the FG coefficients (\ref{real FG coefficients}). By computing the variation of the perturbed one-point correlator $\langle{O}\rangle^{[1]}=2(\Phi^{[1]}_{(2)}+\Psi^{[1]}_{(2)})$, we recover the two-point correlator (\ref{scalar two-point correlator}).\par
In computing higher-order coefficients, the back-reaction of the scalar field to the bulk geometry needs to be considered. The $k$-th order EOM (for $k\geq 2$) is a  nonhomogeneous differential equation given by
\begin{align}
    \frac{1}{\sqrt{\mathcal{G}^{[0]}}}\partial_{\mu}\Big(\sqrt{\mathcal{G}^{[0]}}\mathcal{G}^{[0]\mu\nu}\partial_{\nu}\Phi^{[k]}\Big)=M^{[k]}. \label{n-th scalar EOM}
\end{align}
Here $M^{[k]}$ is a functional of $f(\varphi,t)$, whose explicit form is written in terms of the lower-order coefficients of the metric and the scalar field,
\begin{align}
    M^{[k]}=-\sum_{l=1}^{k-1}\Big[\frac{1}{\sqrt{\mathcal{G}}}\partial_{\mu}\Big(\sqrt{\mathcal{G}}\mathcal{G}^{\mu\nu}\partial_{\nu}\Big)\Big]^{[k-l]}\Phi^{[l]}. \label{M's definition}
\end{align}
Plugging (\ref{metric thermal AdS3})(\ref{Fourier expansion})(\ref{transformation}) into (\ref{n-th scalar EOM}), we obtain
\begin{align}
    x(1-x)\frac{d^2}{dx^2}y^{[k]}_{m,n}+[\gamma-(\alpha+\beta+1)x]\frac{d}{dx}y^{[k]}_{m,n}-\alpha\beta y^{[k]}_{m,n}=\tilde{M}^{[k]}_{m,n},\label{yk's equation}
\end{align}
where $\tilde{M}^{[k]}_{m,n}=4(1-x)^2x^{\frac{n}{2}}M^{[k]}_{m,n}$ and $M^{[k]}_{m,n}$ is the Fourier coefficient of $M^{[k]}$. Consider a Green's function with the following Dirichlet boundary problem,
\begin{align}
    &\Big(x(1-x)\partial_x^2+[\gamma-(\alpha+\beta+1)x]\partial_x-\alpha\beta\Big)\mathscr{G}_{m,n}(x,x_0)=\delta(x-x_0),\notag\\
    &(1-x)\mathscr{G}_{m,n}(x,x_0)\Big|_{x\to 1}=0,\ \ \ \ \ \ \ x^{\frac{n}{2}}\mathscr{G}_{m,n}(x,x_0)\Big|_{x\to 0}\ \ \ \text{regular}.
\end{align}
The explicit form of $\mathscr{G}_{m,n}$ is discussed in appendix \ref{appendix Green's function 2}. $y^{[k]}_{m,n}$ can be formally written as
\begin{align}
    y^{[k]}_{m,n}(x)=&4\int_0^1dx_0(1-x_0)^2x_0^{\frac{n}{2}}{M}^{[k]}_{m,n}(x_0)\mathscr{G}_{m,n}(x,x_0). \label{ykm,n}
\end{align}
To obtain the exact solution, we need to calculate $M^{[k]}_{m,n}(x_0)$, which contains the metric coefficients $\mathcal{G}^{[l]}_{\mu\nu}$ for $l\leq k-1$.  The first-order perturbation $\mathcal{G}^{[1]}_{\mu\nu}=0$ due to the absence of matter terms in the first-order of (\ref{Einstein's equations 0}). Consequently, $M^{[2]}=0$, resulting in the vanishing of the connected three-point correlator. In fact, since the bulk scalar theory is free, any higher-point scalar correlator\footnote{Here we are referring to a correlator that has more than two scalar operator insertions.} at order $\mathcal{O}(G^0)$ can be factorized into a product of lower-point scalar correlators. Therefore, the connected part of the correlator can only be non-zero at higher orders of $G$. To see this, we consider the connected four-point scalar correlator, which corresponds to $M^{[3]}$ in (\ref{M's definition}). Expanding the Einstein's equation (\ref{Einstein's equations 0}) to the second-order in $\epsilon$, we obtain
\begin{align}
    &\frac{1}{2}[\nabla_{\mu}\nabla^{\sigma}\mathcal{G}^{[2]}_{\sigma\nu}+\nabla_{\nu}\nabla^{\sigma}\mathcal{G}^{[2]}_{\sigma\mu}-\nabla_{\mu}\nabla_{\nu}\mathcal{G}^{[2]\sigma}_{\sigma}-\nabla^2\mathcal{G}^{[2]}_{\mu\nu}]-2\mathcal{G}^{[2]}_{\mu\nu}=-8\pi G\partial_{\mu}\Phi^{[1]}\partial_{\nu}\Phi^{[1]}, \label{equation of G2}
\end{align}
where $\nabla_{\mu}$ is the covariant derivative operator of the metric $\mathcal{G}^{[0]}_{\mu\nu}$, and Greek indices are raised by $\mathcal{G}^{[0]\mu\nu}$. The second-order perturbation $\mathcal{G}^{[2]}_{\mu\nu}$ is proportional to Newton’s constant $G$, yielding a non-zero connected four-point scalar correlator at order $\mathcal{O}(G^1)$, which is suppressed in the semiclassical limit. Unfortunately, due to the lack of a viable method for solving the differential equation (\ref{equation of G2}), we are unable to provide an explicit form of the four-point correlator in this context. Similarly, one can expand Einstein's equation to higher orders in $\epsilon$ and obtain differential equations for higher order perturbations of the metric. One can verify that these perturbations are always proportional to $G$, implying that the connected higher-point scalar correlators are non-zero at order $\mathcal{O}(G^1)$.\par
As a direct generalization, we can activate the scalar field source $\phi_{(0)}(z)=\epsilon(z)$. The boundary field theory should then be considered as the CFT deformed by the marginal operator $O$,
\begin{align}
    S'=&S_{CFT}+\int_{\partial\mathcal{M}}d^2z\epsilon(z)O(z).
\end{align}
The deformed one-point correlators can be perturbatively constructed from the correlators on the pure gravitational saddle,
\begin{align}
    \langle{T_{ij}(z)}\rangle_{\epsilon}=&\langle{T_{ij}}\rangle+\int_{\partial\mathcal{M}}d^2z_1\int_{\partial\mathcal{M}}d^2z_2\epsilon(z_1)\epsilon(z_2)\langle{T_{ij}(z)O(z_1)O(z_2)}\rangle+O(\epsilon^3),\notag\\
    \langle{O(z)}\rangle_{\epsilon}=&\int_{\partial\mathcal{M}}d^2z_1\epsilon(z_1)\langle{O(z)O(z_1)}\rangle+O(\epsilon^2).
\end{align}

\subsubsection{Two-point and three-point mixed correlators}

In terms of holographic correlators (\ref{holographic correlator}), we specify the boundary values of bulk fields as
\begin{align}
        \phi_{(0)}(z)=&\epsilon_1 f(z),\notag\\
    g_{(0)ij}(z)=&\eta_{ij}+\epsilon_2\chi_{ij}(z),\label{perturbed boundary values}
\end{align}
where $\epsilon_1$ and $\epsilon_2$ are infinitesimal parameters. The perturbed one-point correlators can be expanded in powers of $\epsilon_1$ and $\epsilon_2$,
\begin{align}
\langle{T_{ij}}\rangle^{[\epsilon_1,\epsilon_2]}=&\sum_{k,l=1}^{\infty}\epsilon_1^k\epsilon_2^l\langle{T_{ij}}\rangle^{[k,l]},\notag\\
\langle{O}\rangle^{[\epsilon_1,\epsilon_2]}=&\sum_{k,l=1}^{\infty}\epsilon_1^k\epsilon_2^l\langle{O}\rangle^{[k,l]}. \label{perturbed one-point correlators}
\end{align}
Plugging (\ref{perturbed boundary values})(\ref{perturbed one-point correlators}) into (\ref{massless trace relation})(\ref{massless conservation equation}), we obtain
\begin{align}
\sum_{m,k,l=0}^{\infty}\epsilon_1^{k}\epsilon_2^{l+m}g^{[0,m]ij}_{(0)}\langle{T_{ij}}\rangle^{[k,l]}=&-\sum_{m=0}^{\infty}\epsilon_2^m\Big[\frac{1}{16\pi G}R^{[0,m]}_{(0)}+\frac{1}{2}\epsilon_1^2g^{[0,m]ij}_{(0)}\partial_if\partial_jf\Big],\notag\\
\sum_{m,k,l=0}^{\infty}\epsilon_1^{k}\epsilon_2^{l+m}\nabla_{(0)}^{[0,m]j}\langle{T_{ij}}\rangle^{[k,l]}=&\sum_{k,l=0}^{\infty}\epsilon_1^{k+1}\epsilon_2^{l}\langle{O(z)}\rangle^{[k,l]}\partial_i f.\label{massless expansion of correlator's equation}
\end{align}
Here $g^{ij}_{(0)}$, $R_{(0)}$, and $\nabla_{(0)}^j$ depend only on the boundary metric $g_{(0)ij}$. We derive the differential equations for stress tensor correlators and mixed correlators from the coefficients of each order in (\ref{massless expansion of correlator's equation}), while the scalar correlators remain unspecified.\par
For instance, the first-order coefficients satisfy
\begin{align}
\eta^{ij}\langle{T_{ij}}\rangle^{[1,0]}=&0,\label{massless 1st order 1}\\
\partial^j\langle{T_{ij}}\rangle^{[1,0]}=&\langle{O}\rangle^{[0,0]}\partial_i f,\label{massless 1st order 2}\\
\eta^{ij}\langle{T_{ij}}\rangle^{[0,1]}=&\frac{1}{16\pi G}[\partial^i\partial^j\chi_{ij}-\partial^i\partial_i\chi^{j}_{j}]+\chi^{ij}\langle{T_{ij}}\rangle^{[0,0]},\label{massless 1st order 3}\\
\partial^{j}\langle{T_{ij}}\rangle^{[0,1]}=&\frac{1}{2}[(2\partial^k\chi^j_{k}-\partial^j\chi^k_{k})\langle{T_{ij}}\rangle^{[0,0]}+\partial_i\chi^{jk }\langle{T_{jk}}\rangle^{[0,0]}],\label{massless 1st order 4}
\end{align}
In deriving (\ref{massless 1st order 4}), we have utilized the translation invariance of $\langle{T_{ij}}\rangle^{[0,0]}$, with indices raised by $\eta^{ij}$. It is convenient to work in the complex coordinates $(z,\bar z)$ with the Euclidean metric $\eta_{z\bar z}=\eta_{\bar zz}=\frac{1}{2}$, $\eta_{zz}=\eta_{\bar z\bar z}=0$. Taking functional derivatives on both sides of (\ref{massless 1st order 1})(\ref{massless 1st order 2})(\ref{massless 1st order 3})(\ref{massless 1st order 4}), we obtain
\begin{align}
\frac{\delta\langle{T_{z\bar z}(z)}\rangle}{\delta \phi_{(0)}(z_1)}=&0,\label{massless 1st order 5}\\
\partial_{\bar z}\frac{\delta\langle{T_{zz}(z)}\rangle}{\delta \phi_{(0)}(z_1)}=&\frac{1}{2}\langle{O}\rangle\partial_z \delta(z-z_1),\label{massless 1st order 6}\\
\partial_{z}\frac{\delta\langle{T_{\bar z\bar z}(z)}\rangle}{\delta \phi_{(0)}(z_1)}=&\text{ c.c. of }\partial_{\bar z}\frac{\delta\langle{T_{zz}(z)}\rangle}{\delta \phi_{(0)}(z_1)},\label{massless 1st order 6bar}\\
\frac{\delta\langle{T_{z\bar z}(z)}\rangle}{\delta g_{(0)ij}(z_1)}=&\frac{1}{16\pi G}\Big(\partial^2_z\frac{\delta\chi_{\bar z\bar z}(z)}{\delta \chi_{ij}(z_1)}-\partial_z\partial_{\bar z}\frac{\delta\chi_{z\bar z}(z)}{\delta \chi_{ij}(z_1)}+\partial^2_{\bar z}\frac{\delta\chi_{zz}(z)}{\delta \chi_{ij}(z_1)}\Big)\notag\\
&+\frac{\delta\chi_{zz}(z)}{\delta \chi_{ij}(z_1)}\langle{T_{\bar z\bar z}}\rangle+\frac{\delta\chi_{z\bar z}(z)}{\delta \chi_{ij}(z_1)}\langle{T_{z\bar z}}\rangle+\frac{\delta\chi_{\bar z\bar z}(z)}{\delta \chi_{ij}(z_1)}\langle{T_{zz}}\rangle,\label{massless 1st order 7}\\
\partial_{\bar z}\frac{\delta\langle{T_{zz}(z)}\rangle}{\delta g_{(0)ij}(z_1)}=&\frac{1}{16\pi G}\Big(-\partial_z\partial_{\bar z}^2 \frac{\delta\chi_{zz}(z)}{\delta \chi_{ij}(z_1)}+\partial^2_z\partial_{\bar z}\frac{\delta\chi_{z\bar z}(z)}{\delta \chi_{ij}(z_1)}-\partial^3_z\frac{\delta\chi_{\bar z\bar z}(z)}{\delta \chi_{ij}(z_1)}\Big)\notag\\
&+2\partial_z\frac{\delta\chi_{\bar z\bar z}(z)}{\delta \chi_{ij}(z_1)}\langle{T_{zz}}\rangle+2\partial_{\bar z}\frac{\delta\chi_{zz}(z)}{\delta \chi_{ij}(z_1)}\langle{T_{z\bar z}}\rangle,\label{massless 1st order 8}\\
\partial_{z}\frac{\delta\langle{T_{\bar z\bar z}(z)}\rangle}{\delta g_{(0)ij}(z_1)}=&\text{ c.c. of }\partial_{\bar z}\frac{\delta\langle{T_{zz}(z)}\rangle}{\delta g_{(0)\bar i\bar j}(z_1)}.\label{massless 1st order 8bar}
\end{align}
The first three equations determine the two-point mixed correlators. (\ref{massless 1st order 5}) gives $\langle{T_{z\bar z}(z)O(z_1)}\rangle=0$. (\ref{massless 1st order 6}) and (\ref{massless 1st order 6bar}) are first-order differential equations, which can be solved as 
\begin{align}
\frac{\delta\langle{T_{zz}(z)}\rangle}{\delta \phi_{(0)}(z_1)}=&\frac{1}{\text{Im}\tau}\int_{\partial\mathcal{M}}d^2w\frac{\delta\langle{T_{zz}(w)}\rangle}{\delta \phi_{(0)}(z_1)}+\frac{1}{2\pi}\langle{O}\rangle\partial_{z}G_{\tau}(z-z_1),\notag\\
\frac{\delta\langle{T_{\bar z\bar z}(z)}\rangle}{\delta \phi_{(0)}(z_1)}=&\frac{1}{\text{Im}\tau}\int_{\partial\mathcal{M}}d^2w\frac{\delta\langle{T_{\bar z\bar z}(w)}\rangle}{\delta \phi_{(0)}(z_1)}+\frac{1}{2\pi}\langle{O}\rangle\partial_{\bar z}\overline{G_{\tau}(z-z_1)}, \label{massless 2pt solution 1}
\end{align}
where $G_{\tau}(z-z_1)$ is the torus Green's function with modular parameter $\tau$ (see appendix \ref{appendix torus Green's function} for the definition), and $\text{Im}\tau$ is the imaginary part of $\tau$. As we discussed above, (\ref{massless 2pt solution 1}) leaves two one-point-averaged correlators undetermined. By definition (\ref{holographic correlator}), we can rewrite them as the global metric variations of $\langle{O}\rangle$, and then apply (\ref{alternative method}) to obtain
\begin{align}
\int_{\partial\mathcal{M}}d^2w\frac{\delta\langle{T_{zz}(w)}\rangle}{\delta \phi_{(0)}(z_1)}=&\frac{1}{2}\int_{\partial\mathcal{M}}d^2w\frac{\delta\langle{O(z_1)}\rangle}{\delta g_{(0)\bar z\bar z}(w)}=-i\text{Im}\tau\partial_\tau\langle{O}\rangle-\frac{1}{2}\langle{O}\rangle,\notag\\
\int_{\partial\mathcal{M}}d^2w\frac{\delta\langle{T_{\bar z\bar z}(w)}\rangle}{\delta \phi_{(0)}(z_1)}=&\frac{1}{2}\int_{\partial\mathcal{M}}d^2w\frac{\delta\langle{O(z_1)}\rangle}{\delta g_{(0)zz}(w)}=i\text{Im}\tau\partial_{\bar\tau}\langle{O}\rangle-\frac{1}{2}\langle{O}\rangle.
\end{align}
Putting everything together, we obtain 
\begin{align}
\langle{T_{zz}(z)O(z_1)}\rangle=&-i\partial_\tau\langle{O}\rangle-\frac{1}{2\pi}[\wp_\tau(z-z_1)+2\zeta_\tau(\frac{1}{2})]\langle{O}\rangle,\notag\\
\langle{T_{\bar z\bar z}(z)O(z_1)}\rangle=&\text{ c.c. of } \langle{T_{zz}(z)O(z_1)}\rangle ,\notag\\
\langle{T_{z\bar z}(z)O(z_1)}\rangle=&0,
\end{align}
where $\wp_\tau(z)$ and $\zeta_\tau(z)$ are the Weierstrass $\wp$-function and $\zeta$-function respectively. Note that the two-point mixed correlators are expressed as functions of the scalar one-point correlator. The latter is determined by the FG coefficients of the saddle (\ref{massless case one-point correlators}). In our context, the scalar one-point correlator vanishes  as a consequence of the $Z_2$ symmetry $\Phi\to -\Phi$. More generally, since our calculations are based on the semiclassical approximation, the precise one-point correlator could be nonzero at higher-orders in $1/c$, but it is suppressed in the large-$c$ limit. The last three equations (\ref{massless 1st order 7})(\ref{massless 1st order 8})(\ref{massless 1st order 8bar}) determine the two-point stress tensor correlators. Note that these equations are derived with the scalar field turned off, and stress tensor correlators obtained from them agree with the results in pure gravity \cite{He:2023hoj}.\par
Let us further consider the second-order coefficients of (\ref{massless expansion of correlator's equation}),
\begin{align}
\eta^{ij}\langle{T_{ij}}\rangle^{[2, 0]}=&-\frac{1}{2}\partial^if\partial_if,\label{massless 2nd order 1}\\
\partial^j\langle{T_{ij}}\rangle^{[2,0]}=&\langle{O}\rangle^{[1.0]}\partial_if,\label{massless 2nd order 2}\\
\eta^{ij}\langle{T_{ij}}\rangle^{[1,1]}=&\chi^{ij}\langle{T_{ij}}\rangle^{[1,0]},\label{massless 2nd order 3}\\
\partial^j\langle{T_{ij}}\rangle^{[1,1]}=&\chi^{jk}\partial_k\langle{T_{ij}}\rangle^{[1,0]}+\frac{1}{2}[(2\partial^k\chi^j_{k}-\partial^j\chi^k_{k})\langle{T_{ij}}\rangle^{[1,0]}+\partial_i\chi^{jk}\langle{T_{jk}}\rangle^{[1,0]}]\notag\\
&+\langle{O}\rangle^{[0,1]}\partial_if,\label{massless 2nd order 4}\\
\eta^{ij}\langle{T_{ij}}\rangle^{[0,2]}=&\chi^{ij}\langle{T_{ij}}\rangle^{[0,1]}-\chi^{ik}\chi^j_k\langle{T_{ij}}\rangle^{[0,0]}-\frac{1}{64\pi G}[4\partial^i\chi_{i}^j\partial^k\chi_{jk}-4\partial^i\chi_{i}^j\partial_j\chi^k_{k}\notag\\
&+\partial^i\chi_{j}^j\partial_i\chi^k_{k}-3\partial^i\chi^j_{k}\partial_i\chi^{k}_j+2\partial^i\chi^{j}_k\partial_j\chi^k_{i}-4\chi^{ij}(\partial_i\partial_j\chi^k_k\notag\\
&+\partial^k\partial_k\chi_{ij}-2\partial_i\partial^k\chi_{jk})],\label{massless 2nd order 5}\\
\partial^j\langle{T_{ij}}\rangle^{[0,2]}=&\chi^{jk}\partial_k\langle{T_{ij}}\rangle^{[0,1]}+\frac{1}{2}[(2\partial^k\chi^j_{k}-\partial^j\chi^k_{k})\langle{T_{ij}}\rangle^{[0,1]}+\partial_i\chi^{jk}\langle{T_{jk}}\rangle^{[0,1]}]\notag\\
&+\frac{1}{2}[\chi^{kl}(\partial^j\chi_{ik}-\partial_i\chi^j_k-\partial_k\chi^j_i)+\chi^{jk}(\partial_k\chi^l_i-\partial_i\chi^l_k-\partial^l\chi_{ik})]\langle{T_{jl}}\rangle^{[0,0]}\notag\\
&+\frac{1}{2}[\chi^{kl}(\partial^j\chi_{kl}-2\partial_k\chi^k_l)+\chi^{jk}(\partial_k\chi^l_l-2\partial^l\chi_{kl})]\langle{T_{ij}}\rangle^{[0,0]}.\label{massless 2nd order 6}
\end{align}
The last two equations (\ref{massless 2nd order 5})(\ref{massless 2nd order 6}) give the three-point stress tensor correlators, which have been considered in our previous work. For the remaining four equations,
(\ref{massless 2nd order 1})(\ref{massless 2nd order 2}) give the correlators of type $\langle{T_{ij}(z)O(z_1)O(z_2)}\rangle$,
\begin{align}
    \langle{T_{zz}(z)O(z_1)O(z_2)}\rangle=&-i\partial_\tau\langle{O(z_1)O(z_2)}\rangle-\frac{1}{2\pi}\sum_{n=1}^2\Big[\Big([\wp_\tau(z-z_n)+2\zeta_\tau(\frac{1}{2})]\notag\\
    &+[\zeta_\tau(z-z_n)-2\zeta_\tau(\frac{1}{2})(z-z_n)]\partial_{z_n}\Big)\langle{O(z_1)O(z_2)}\rangle\notag\\
    &-\Big(\pi\partial_{z_n}\delta(z-z_n)\partial_{z_n}+[\wp_\tau(z-z_n)+2\zeta_\tau(\frac{1}{2})]\partial_{z_n}\partial_{\bar z_n}\Big)\delta(z_1-z_2)\Big],\notag\\
    \langle{T_{\bar z\bar z}(z)O(z_1)O(z_2)}\rangle=&\text{ c.c. of }\langle{T_{zz}(z)O(z_1)O(z_2)}\rangle,\notag\\
    \langle{T_{z\bar z}(z)O(z_1)O(z_2)}\rangle=&-\frac{1}{2}\partial_{z}\delta(z-z_1)\partial_{\bar z}\delta(z-z_2)+(z_1\leftrightarrow z_2),\label{massless three-point 1}
\end{align}
and (\ref{massless 2nd order 3})(\ref{massless 2nd order 4}) give the correlators of type $\langle{T_{ij}(z)T_{kl}(z_1)O(z_2)}\rangle$,
\begin{align}
    \langle{T_{zz}(z)T_{zz}(z_1)O(z_2)}\rangle=&-i\partial_\tau\langle{T_{zz}(z_1)O(z_2)}\rangle-\frac{1}{2\pi}\Big[2\Big(\wp_\tau(z-z_1)+2\zeta_\tau(\frac{1}{2})\Big)\notag\\
    &+\Big(\wp_\tau(z-z_2)+2\zeta_\tau(\frac{1}{2})\Big)+\Big(\zeta_\tau(z-z_1)-2\zeta_\tau(\frac{1}{2})(z-z_1)\Big)\partial_{z_1}\notag\\
    &+\Big(\zeta_\tau(z-z_2)-2\zeta_\tau(\frac{1}{2})(z-z_1)\Big)\partial_{z_2}\Big]\langle{T_{zz}(z_1)O(z_2)}\rangle,\notag\\
    \langle{T_{\bar z\bar z}(z)T_{\bar z\bar z}(z_1)O(z_2)}\rangle=&\text{ c.c. of } \langle{T_{zz}(z)T_{zz}(z_1)O(z_2)}\rangle,\notag\\
    \langle{T_{zz}(z)T_{\bar z\bar z}(z_1)O(z_2)}\rangle=&-i\partial_\tau\langle{T_{\bar z\bar z}(z_1)O(z_2)}\rangle-\frac{1}{2\pi}\Big[\Big(\wp_\tau(z-z_2)+2\zeta_\tau(\frac{1}{2})\Big)\notag\\
    &+\Big(\zeta_\tau(z-z_2)-2\zeta_\tau(\frac{1}{2})(z-z_1)\Big)\partial_{z_2}\Big]\langle{T_{\bar z\bar z}(z_1)O(z_2)}\rangle,\notag\\
    \langle{T_{z\bar z}(z)T_{zz}(z_1)O(z_2)}\rangle=&\frac{1}{2}\delta(z-z_1)\langle{T_{zz}(z)O(z_2)}\rangle,\notag\\
    \langle{T_{z\bar z}(z)T_{\bar z\bar z}(z_1)O(z_2)}\rangle=&\text{ c.c. of }\langle{T_{z\bar z}(z)T_{zz}(z_1)O(z_2)}\rangle,\notag\\
    \langle{T_{z\bar z}(z)T_{z\bar z}(z_1)O(z_2)}\rangle=&0. \label{massless three-point 2}
\end{align}
In this context, we have utilized the principle of translation invariance, represented as $\sum_{i=1}^N\partial_{z_i}\langle{X(z_1,...,z_N)}\rangle=0$. While our calculations effectively reduce the stress tensors in the mixed correlators, they leave the scalar correlators undetermined. Unfortunately, the application of FG expansions to the scalar equations of motion (EOM) results in the breakdown of the constraint on $\phi_{(2)}$, rendering it impossible to establish a differential equation for $\langle{O}\rangle$. Determining scalar correlators requires a global solution to the scalar EOM, a topic we have explored in section \ref{3.1.2}.
\subsubsection{Recurrence relations of mixed correlators}\label{3.1.1}
In principle, with increasingly tedious calculations, one can obtain the arbitrary higher-point mixed correlator by the above method. Some special perturbations simplify (\ref{massless expansion of correlator's equation}) and lead to recurrence relations. For instance, if we only turn on the variations of the scalar field $\delta\phi_{(0)}(z)=\epsilon_1 f(z)$ and $\bar z\bar z$ component of the metric $\delta g_{(0)\bar z\bar z}(z)=\epsilon_2 F(z)$, geometric quantities can be represented as polynomials in $\epsilon_2$,
\begin{align}
    g^{ij}_{(0)}=&\eta^{ij}-4\epsilon_2 F\delta^i_z\delta^j_z,\notag\\
    \Gamma^{k}_{(0)ij}=&\epsilon_2[\delta^k_z\delta^{\bar z}_i\partial_jF+\delta^k_z\delta^{\bar z}_j\partial_iF-\frac{1}{2}\eta^{kl}\delta^{\bar z}_i\delta^{\bar z}_j\partial_lF]+2\epsilon_2^2\delta^k_z\delta^{\bar z}_i\delta^{\bar z}_jF\partial_zF,\notag\\
    R_{(0)}=&-4\epsilon_2\partial_z^2F.\label{Recurrence relations geometric quantities}
\end{align}
Plugging (\ref{Recurrence relations geometric quantities}) into (\ref{massless expansion of correlator's equation}) we obtain
\begin{align}
\langle{T_{z\bar z}}\rangle^{[k,l]}=&F\langle{T_{zz}}\rangle^{[k,l-1]}+\frac{\delta_{k,0}\delta_{l,1}}{16\pi G}\partial_z^2F-\frac{\delta_{k,2}}{2}[\delta_{l,0}\partial_zf\partial_{\bar z}f-\delta_{l,1}(\partial_zf)^2F],\notag\\
\partial_{\bar z}\langle{T_{zz}}\rangle^{[k,l]}=&\frac{1}{2}\langle{O}\rangle^{[k-1,l]}\partial_zf+2F\partial_z\langle{T_{zz}}\rangle^{[k,l-1]}+3\partial_zF\langle{T_{zz}}\rangle^{[k,l-1]}-\partial_z\langle{T_{z\bar z}}\rangle^{[k,l]}, \label{massless recurrence relations 0}
\end{align}
where $\delta_{m,n}$ is the Kronecker symbol defined as $\delta_{m,n}=1$ for $m=n$ and $\delta_{m,n}=0$ for $m\neq n$. The terms containing the Kronecker symbols come from the conformal anomaly and contribute to some specific correlators (as contact terms most of the time). Taking functional derivatives on both sides of (\ref{massless recurrence relations 0}) and solving the differential equation, we obtain
\begin{align}
    &\frac{\delta^{k+l}\langle{T_{zz}(z)}\rangle}{\prod_{i=1}^k\delta\phi_{(0)}(x_i)\prod_{j=1}^l\delta g_{(0)\bar z\bar z}(y_j)}\notag\\
    =&C(x,y)+\frac{1}{2\pi}\sum_{i=1}^k\Big[\Big(\partial_zG_\tau(z-x_i)-G_\tau(z-x_i)\partial_{x_i}\Big)\frac{\delta^{k+l-1}\langle{O(x_i)}\rangle}{\prod_{i'\neq i}\delta\phi_{(0)}(x_{i'})\prod_{j=1}^l\delta g_{(0)\bar z\bar z}(y_j)}\Big]\notag\\
    &+\frac{1}{\pi}\sum_{j=1}^l\Big[\Big(2\partial_zG_\tau(z-y_j)-G_\tau(z-y_j)\partial_{y_j}\Big)\frac{\delta^{k+l-1}\langle{T_{zz}(y_j)}\rangle}{\prod_{i=1}^k\delta\phi_{(0)}(x_{i})\prod_{j'\neq j}\delta g_{(0)\bar z\bar z}(y_{j'})}\Big]\notag\\
    &-\frac{\delta_{k,0}\delta_{l,1}}{16\pi^2G}\partial_z^3G_\tau(z-y)+\frac{\delta_{k,2}}{2\pi}\Big[\delta_{l,0}\partial_{\bar x_1}[\partial_{x_1}G_\tau(z-x_1)\partial_{x_1}\delta(x_1-x_2)]\notag\\
    &+\delta_{l,1}\partial_yG_\tau(z-y)\partial_y\delta(y-x_1)\partial_y\delta(y-x_2)+(x_1\leftrightarrow x_2)\Big].\label{massless recurrence relations 1}
\end{align}
The undetermined function $C(x,y)=\frac{1}{\text{Im}\tau}\int d^2w\frac{\delta^{k+l}\langle{T_{zz}(w)}\rangle}{\prod_{i=1}^k\delta\phi_{(0)}(x_i)\prod_{j=1}^l\delta g_{(0)\bar z\bar z}(y_j)}$ is calculated using (\ref{holographic correlator})(\ref{alternative method})(\ref{massless recurrence relations 0}),
\begin{align}
    &C(x,y)\notag\\
    =&-\frac{2^{l}}{2\text{Im}\tau}\Big[2i\text{Im}\tau\partial_\tau+(k+2l)+\sum_{i=1}^k(x_i-\bar x_i)\partial_{x_i}+\sum_{j=1}^l(y_j-\bar y_j)\partial_{y_j}\Big]\langle{\prod_{i=1}^kO(x_i)\prod_{j=1}^lT_{zz}(y_j)}\rangle\notag\\
    &+\frac{\delta_{k,2}}{2\text{Im}\tau}\Big[\delta_{l,0}\partial_{x_1}\partial_{\bar x_1}\delta(x_1-x_2)+\delta_{l,1}\partial_y\delta(y-x_1)\partial_y\delta(y-x_2)+(x_1\leftrightarrow x_2)\Big].\label{massless recurrence relations for C}
\end{align}
Combining (\ref{massless recurrence relations 1}) and (\ref{massless recurrence relations for C}), we find the recurrence relation
\begin{align}
    &\langle{T_{zz}(z)\prod_{i=1}^kO(x_i)\prod_{j=1}^lT_{zz}(y_j)}\rangle\notag\\
    =&-\frac{1}{2\pi}\Big[2\pi i\partial_\tau+\sum_{i=1}^k\Big(\wp_\tau(z-x_i)+2\zeta_\tau(\frac{1}{2})+[\zeta_\tau(z-x_i)-2\zeta_\tau(\frac{1}{2})(z-x_i)]\partial_{x_i}\Big)\notag\\
    &+\sum_{j=1}^l\Big(2[\wp_\tau(z-y_j)+2\zeta_\tau(\frac{1}{2})]+[\zeta_\tau(z-y_j)-2\zeta_\tau(\frac{1}{2})(z-y_j)]\partial_{y_j}\Big)\Big]\langle{\prod_{i=1}^kO(x_i)\prod_{j=1}^lT_{zz}(y_j)}\rangle\notag\\
    &+\frac{\delta_{k,0}\delta_{l,1}}{32\pi^2 G}\wp''_\tau(z-y)+\frac{\delta_{k,2}}{2\pi}\Big[\delta_{l,0}\Big(\pi\partial_{x_1}\delta(z-x_1)\partial_{x_1}+[\wp_\tau(z-x_1)+2\zeta_\tau(\frac{1}{2})]\partial_{x_1}\partial_{\bar x_1}\Big)\delta(x_1-x_2)\notag\\
    &+\frac{\delta_{l,1}}{2}[\wp_\tau(z-y)+2\zeta_\tau(\frac{1}{2})]\partial_y\delta(y-x_1)\partial_y\delta(y-x_2)+(x_1\leftrightarrow x_2)\Big].\label{massless recurrence relations 2}
\end{align}
Note that (\ref{massless recurrence relations 0}) should also contain a differential equation for $\langle{T_{\bar z\bar z}}\rangle^{[k,l]}$,
\begin{align}
    \partial_z\langle{T_{\bar z\bar z}}\rangle^{[k,l]}=&\frac{1}{2}\langle{O}\rangle^{[k-1,l]}\partial_{\bar z}f+\partial_{\bar z}F\langle{T_{zz}}\rangle^{[k,l-1]}+2\partial_z[F\langle{T_{z\bar z}}\rangle^{[k,l-1]}]-\partial_{\bar z}\langle{T_{z\bar z}}\rangle^{[k,l]}.
\end{align}
Through a similar procedure, we obtain
\begin{align}
    &\langle{T_{\bar z\bar z}(z)\prod_{i=1}^kO(x_i)\prod_{j=1}^lT_{zz}(y_j)}\rangle\notag\\
    =&-\frac{1}{2\pi}\Big[-2\pi i\partial_{\bar\tau}+\sum_{i=1}^k\Big(\overline{\wp_\tau(z-x_i)}+2\overline{\zeta_\tau(\frac{1}{2})}+[\overline{\zeta_\tau(z-x_i)}-2\overline{\zeta_\tau(\frac{1}{2})}(\bar z-\bar x_i)]\partial_{\bar x_i}\Big)\notag\\
    &+\sum_{j=1}^l\Big(\overline{\zeta_\tau(z-y_j)}-2\overline{\zeta_\tau(\frac{1}{2})}(\bar z-\bar y_j)\Big)\partial_{\bar y_j}\Big]\langle{\prod_{i=1}^kO(x_i)\prod_{j=1}^lT_{zz}(y_j)}\rangle\notag\\
    &+\sum_{j=1}^l\delta(z-y_j)\langle{T_{z\bar z}(z)\prod_{i=1}^kO(x_i)\prod_{j'\neq j}T_{zz}(y_{j'})}\rangle-\frac{\delta_{k,0}\delta_{l,1}}{32\pi G}\partial_z\partial_{\bar z}\delta(z-y)\notag\\
    &+\frac{\delta_{k,2}}{2\pi}\Big[\delta_{l,0}\Big(\pi\partial_{\bar x_1}\delta(z-x_1)\partial_{\bar x_1}+[\overline{\wp_\tau(z-x_1)}+2\overline{\zeta_\tau(\frac{1}{2})}]\partial_{x_1}\partial_{\bar x_1}\Big)\delta(x_1-x_2)\notag\\
    &+\frac{\delta_{l,1}}{2}[\overline{\wp_\tau(z-y)}+2\overline{\zeta_\tau(\frac{1}{2})}]\partial_y\delta(y-x_1)\partial_y\delta(y-x_2)+(x_1\leftrightarrow x_2)\Big].\label{massless recurrence relations 3}
\end{align}
With the omission of the contact terms, (\ref{massless recurrence relations 2}) and (\ref{massless recurrence relations 3}) agree with the Ward identity in the torus CFT \cite{Felder:1989vx, Chang:1991ht, DiFrancesco:1997nk, He:2020udl} by the Brown-Henneaux central charge relation $c=\frac{3}{2G}$ \cite{Brown:1986nw}. The dual operator $O$ is a primary operator with conformal weights $h=\bar h=1$.


\subsection{Massive scalar field}\label{3.2}
We further compute mixed correlators in the case of gravity coupled to a massive scalar field. Dual operators are classified according to their scaling dimensions. The operator is irrelevant for $\Delta>2$, in which case, we can only consider the infinitesimal source; otherwise, it would cause a self-consistent problem \cite{Witten:1998qj}. For $\Delta=2$, the operator is marginal and has been discussed in section \ref{3.1}. In this section, we are interested in the relevant operator where $\Delta<2$. The scaling dimension has a lower bound $\Delta\geq 1$ called the BF bound \cite{Breitenlohner:1982bm, Breitenlohner:1982jf}, which guarantees the stability of the AdS background.\par
We continue to adopt the FG coordinate system (\ref{Fefferman-Graham coordinates}). Plugging (\ref{Fefferman-Graham coordinates}) into (\ref{Einstein's equations 0})(\ref{scalar EOM}), we obtain the EOMs of the bulk fields,
\begin{align}
    &\rho(2\partial_\rho^2g_{ij}-2\partial_\rho g_{ik}g^{kl}\partial_\rho g_{lj}+g^{kl}\partial_\rho g_{kl}\partial_\rho g_{ij})+R_{(g)ij}-g^{kl}\partial_\rho g_{kl}g_{ij}\notag\\
    &+8\pi G\rho^{1-\Delta}[\Delta(\Delta-2)g_{ij}\phi^2+\rho\partial_i\phi\partial_j\phi]=0,\notag\\
    &\nabla_{(g)i}(g^{kl}\partial_\rho g_{kl})-\nabla_{(g)}^k(\partial_\rho g_{ki})+16\pi G\rho^{1-\Delta}[\frac{2-\Delta}{2}\phi\partial_i\phi+\rho\partial_\rho\phi\partial_i\phi]=0,\notag\\
    &g^{kl}\partial^2_\rho g_{kl}-\frac{1}{2}g^{ij}\partial_\rho g_{jk}g^{kl}\partial_\rho g_{li}+16\pi G\rho^{-\Delta}[(2-\Delta)(\frac{(1-\Delta)}{2}\phi^2+\rho\phi\partial_\rho\phi)+\rho^2\partial_\rho\phi\partial_\rho\phi]=0,\notag\\
    &\Box_{(g)}\phi+2\rho(\partial_{\rho}\text{ln}g\partial_{\rho}\phi+2\partial_{\rho}^2\phi)+(2-\Delta)(\partial_{\rho}\text{ln}g\phi+4\partial_{\rho}\phi)=0, \label{massive Einstein's equations and scalar EOM}
\end{align}
where $\phi(\rho,x)=\rho^{\frac{\Delta}{2}-1}\Phi(\rho,x)$. It is natural to expand $g_{ij}$ and $\phi$ in powers of $\rho$ and then solve the above equations order by order. The bulk EOMs contain $\rho^{-\Delta}$,  and $\Delta$ is not necessarily an integer. We thus cannot simply assume that $g_{ij}$ and $\phi$ can be expanded as formal Taylor series in $\rho$. The authors of \cite{Hung:2011ta} prove that for rational scaling dimensions, the FG expansions include fractional orders of $\rho$. Suppose that $\frac{2-\Delta}{2}=\frac{N}{M}$, where $M$ and $N$ are relatively prime and $0<\frac{N}{M}\leq\frac{1}{2}$. 
As we mentioned in section \ref{2}, each bulk field has two homogeneous solutions, and their expansions in $\rho$ are given by \cite{Hung:2011ta},
\begin{align}
    g^{\pm}_{ij}(\rho,x)=&\sum_{n=0}^{N-1}\rho^n\Big[g^{\pm}_{(2n)ij}+\sum_{m=2}^{\infty}\rho^{m\frac{N}{M}}g^{\pm}_{(2n+2m\frac{N}{M})ij}\Big](x),\notag\\
    \phi^{\pm}(\rho,x)=&\sum_{n=0}^{N-1}\rho^n\Big[\phi^{\pm}_{(2n)}+\sum_{m=1}^{\infty}\rho^{m\frac{N}{M}}\phi^{\pm}_{(2n+2m\frac{N}{M})}\Big](x).\label{massive two homogeneous solutions}
\end{align}
Plugging (\ref{massive two homogeneous solutions}) into (\ref{two homogeneous solutions}) and adding logarithmic terms to the overlap of two solutions, we obtain the FG expansions for $g_{ij}$ and $\phi$, which will be shown later.\par
Let us first assume that $N\geq 2$. Since $M$ and $N$ are relatively prime, the smallest integer that $m\frac{N}{M}$ passes through is $N$. Thus, the first logarithmic term that appears in the scalar field expansion\footnote{The discussion here is slightly different from \cite{Hung:2011ta}. In the latter, the FG expansion of $\phi$ contains $\rho^{1-2\frac{N}{M}}\text{ln}\rho\psi_{(2-4\frac{N}{M})}$ and $\rho^{1-\frac{N}{M}}\text{ln}\rho\psi_{(2-2\frac{N}{M})}$. We can keep these two terms, plug the expansions of $g_{ij}$ and $\phi$ into (\ref{massive Einstein's equations and scalar EOM}), and solve the equations order by order. Eventually we will find that $\psi_{(2-4\frac{N}{M})}=\psi_{(2-2\frac{N}{M})}=0$ for $N\geq 2$.} is $\rho\text{ln}\rho\psi_{(2)}$ instead of $\rho^{1-2\frac{N}{M}}\text{ln}\rho\psi_{(2-4\frac{N}{M})}$. The FG expansions take the forms
\begin{align}
g_{ij}(\rho,x)=&g_{(0)ij}(x)+\sum_{m=2}^{\lfloor\frac{M}{N}\rfloor}\rho^{m\frac{N}{M}}g_{(2m\frac{N}{M})ij}(x)+\rho[g_{(2)ij}+\text{ln}\rho h_{(2)ij}](x)+\cdots,\notag\\
\phi(\rho,x)=&\phi_{(0)}(x)+\sum_{m=1}^{\lfloor\frac{M}{N}\rfloor}\rho^{m\frac{N}{M}}\phi_{(2m\frac{N}{M})}(x)+\rho^{1-2\frac{N}{M}}\phi_{(2-4\frac{N}{M})}(x)+\rho^{1-\frac{N}{M}}\phi_{(2-2\frac{N}{M})}(x)\notag\\
&+\rho[\phi_{(2)}+\text{ln}\rho\psi_{(2)}](x)+\cdots. \label{N>1 Fefferman-Graham expansions}
\end{align}
Plugging (\ref{N>1 Fefferman-Graham expansions}) into (\ref{massive Einstein's equations and scalar EOM}), we obtain
\begin{align}
    h_{(2)ij}=&0,\notag\\
    g^{ij}_{(0)}g_{(2)ij}=&\frac{1}{2}R_{(0)}-64\pi G\frac{N}{M}(1-\frac{N}{M})\phi_{(0)}\phi_{(2-4\frac{N}{M})},\notag\\
    \nabla^{j}g_{(2)ij}=&\partial_{i}(g^{kl}_{(0)}g_{(2)kl})+16\pi G[\frac{N}{M}\phi_{(0)}\partial_i\phi_{(2-4\frac{N}{M})}+(1-\frac{N}{M})\phi_{(2-4\frac{N}{M})}\partial_i\phi_{(0)}]. \label{N>1 coefficients}
\end{align}
The next step is to calculate the one-point correlators $\langle{T_{ij}}\rangle$ and $\langle{O}\rangle$ by taking variations of the renormalized on-shell action. Following the spirit of \cite{deHaro:2000vlm}, we construct the counterterm as (see appendix \ref{appendix Holo Ren} for a detailed derivation)
\begin{align}
 I_{\text{ct}}=&\frac{1}{8\pi G}\int_{\partial\mathcal{M}_{\varepsilon}}d^2x\sqrt{\gamma}\Big[1+\frac{N}{M}\frac{8\pi G\phi^2}{\varepsilon^{-2\frac{N}{M}}}+\sum_{m=3}^{\lfloor{\frac{M}{N}}\rfloor}\frac{(\pi G)^{\frac{m}{2}}\tilde C_{(2m\frac{N}{M})}\phi^m}{\varepsilon^{-m\frac{N}{M}}}+\frac{1}{4}R_{(\gamma)}\text{ln}\varepsilon\Big], \label{N>1 counterterm}
\end{align}
where $\lbrace{\tilde C_{(2m\frac{N}{M})}}\rbrace$ are some constants that do not show up in the finite part of the renormalized on-shell action. The renormalized one-point correlators are given by
\begin{align}
    \langle{T_{ij}}\rangle=&\frac{1}{8\pi G}\Big(g_{(2)ij}-\text{Tr}g_{(2)}g_{(0)ij}\Big)-2\frac{N}{M}\phi_{(0)}\phi_{(2-4\frac{N}{M})}g_{(0)ij},\notag\\
    \langle{O}\rangle=&2(1-2\frac{N}{M})\phi_{(2-4\frac{N}{M})}. \label{N>1 renormalized one-point correlators}
\end{align}
Combining (\ref{N>1 coefficients}) and (\ref{N>1 renormalized one-point correlators}), we obtain the trace relation and the conservation equation,
\begin{align}
    \langle{T^i_{i}}\rangle=&-\frac{1}{16\pi G}R_{(0)}+2\frac{N}{M}\phi_{(0)}\langle{O}\rangle,\notag\\
    \nabla^j_{(0)}\langle{T_{ij}}\rangle=&\langle{O}\rangle\partial_i\phi_{(0)}. \label{N>1 trace relation and conservation equation}
\end{align}
One can compute holographic mixed correlators from (\ref{N>1 trace relation and conservation equation}) by utilizing the same techniques in section \ref{3.1}. In particular, we are concerned with the holographic recurrence relations and wish to compare them with the results in CFT$_2$. Plugging the Taylor expansions (\ref{perturbed one-point correlators}) into (\ref{N>1 trace relation and conservation equation}), we read off 
\begin{align}
\langle{T_{z\bar z}}\rangle^{[k,l]}=&F\langle{T_{zz}}\rangle^{[k,l-1]}+\frac{N}{2M}\langle{O}\rangle^{[k-1,l]}f-\frac{\delta_{k,0}\delta_{l,1}}{16\pi G}\partial_z^2F,\label{N>1 recurrence relation 1}\\
\partial_{\bar z}\langle{T_{zz}}\rangle^{[k,l]}=&\frac{1}{2}\langle{O}\rangle^{[k-1,l]}\partial_zf+2F\partial_z\langle{T_{zz}}\rangle^{[k,l-1]}+3\partial_zF\langle{T_{zz}}\rangle^{[k,l-1]}-\partial_z\langle{T_{z\bar z}}\rangle^{[k,l]},\label{N>1 recurrence relation 2}\\
\partial_z\langle{T_{\bar z\bar z}}\rangle^{[k,l]}=&\frac{1}{2}\langle{O}\rangle^{[k-1,l]}\partial_{\bar z}f+\partial_{\bar z}F\langle{T_{zz}}\rangle^{[k,l-1]}+2\partial_z[F\langle{T_{z\bar z}}\rangle^{[k,l-1]}]-\partial_{\bar z}\langle{T_{z\bar z}}\rangle^{[k,l]}.\label{N>1 recurrence relation 3}
\end{align}
After taking the functional derivatives, (\ref{N>1 recurrence relation 1}) gives $\langle{T_{z\bar z}O...OT_{zz}...T_{zz}}\rangle$, while (\ref{N>1 recurrence relation 2}) and (\ref{N>1 recurrence relation 3}) provide the differential equations of $\langle{T_{zz}O...OT_{zz}...T_{zz}}\rangle$ and $\langle{T_{\bar z\bar z}O...OT_{zz}...T_{zz}}\rangle$, respectively. We solve these differential equations by the torus Green's function and compute the one-point-averaged correlators using (\ref{alternative method}). The final results are listed as follows:
\begin{align}
        &\langle{T_{zz}(z)\prod_{i=1}^kO(x_i)\prod_{j=1}^lT_{zz}(y_j)}\rangle\notag\\
    =&-\frac{1}{2\pi}\Big[2\pi i\partial_\tau+\sum_{i=1}^k\Big((1-\frac{N}{M})[\wp_\tau(z-x_i)+2\zeta_\tau(\frac{1}{2})]+[\zeta_\tau(z-x_i)-2\zeta_\tau(\frac{1}{2})(z-x_i)]\partial_{x_i}\Big)\notag\\
    &+\sum_{j=1}^l\Big(2[\wp_\tau(z-y_j)+2\zeta_\tau(\frac{1}{2})]+[\zeta_\tau(z-y_j)-2\zeta_\tau(\frac{1}{2})(z-y_j)]\partial_{y_j}\Big)\Big]\langle{\prod_{i=1}^kO(x_i)\prod_{j=1}^lT_{zz}(y_j)}\rangle\notag\\
    &+\frac{\delta_{k,0}\delta_{l,1}}{32\pi^2 G}\wp''_\tau(z-y),\label{N>1 recurrence relation 4}\\
    &\langle{T_{\bar z\bar z}(z)\prod_{i=1}^kO(x_i)\prod_{j=1}^lT_{zz}(y_j)}\rangle\notag\\
    =&-\frac{1}{2\pi}\Big[-2\pi i\partial_{\bar\tau}+\sum_{i=1}^k\Big((1-\frac{N}{M})[\overline{\wp_\tau(z-x_i)}+2\overline{\zeta_\tau(\frac{1}{2})}]+[\overline{\zeta_\tau(z-x_i)}-2\overline{\zeta_\tau(\frac{1}{2})}(\bar z-\bar x_i)]\partial_{\bar x_i}\Big)\notag\\
    &+\sum_{j=1}^l\Big(\overline{\zeta_\tau(z-y_j)}-2\overline{\zeta_\tau(\frac{1}{2})}(\bar z-\bar y_j)\Big)\partial_{\bar y_j}\Big]\langle{\prod_{i=1}^kO(x_i)\prod_{j=1}^lT_{zz}(y_j)}\rangle\notag\\
    &+\sum_{j=1}^l\delta(z-y_j)\langle{T_{z\bar z}(z)\prod_{i=1}^kO(x_i)\prod_{j'\neq j}T_{zz}(y_{j'})}\rangle-\frac{\delta_{k,0}\delta_{l,1}}{32\pi G}\partial_z\partial_{\bar z}\delta(z-y).\label{N>1 recurrence relation 5}
\end{align}
Except for the contact terms, (\ref{N>1 recurrence relation 4}) and (\ref{N>1 recurrence relation 5}) agree with the Ward identity in the torus CFT, and $O$ is a primary operator with conformal weights $h=\bar h=1-\frac{N}{M}=\frac{\Delta}{2}$.\par
For the case of $N=1$, the fractional order $\frac{m}{M}$ passes through $1$ and the scalar conformal anomaly $\psi_{(2-\frac{4}{M})}$ emerges. The FG expansions take the forms
\begin{align}
g_{ij}(\rho,x)=&g_{(0)ij}(x)+\sum_{m=2}^{M-1}\rho^{\frac{m}{M}}g_{(\frac{2m}{M})ij}(x)+\rho [g_{(2)ij}+\text{ln}\rho h_{(2)ij}](x)+\cdots,\notag\\
\phi(\rho,x)=&\phi_{(0)}(x)+\sum_{m=1}^{M-3}\rho^{\frac{m}{M}}\phi_{(\frac{2m}{M})}(x)+\rho^{1-\frac{2}{M}}[\phi_{(2-\frac{4}{M})}+\text{ln}\rho\psi_{(2-\frac{4}{M})}](x)+\cdots. \label{N=1 Fefferman-Graham expansions}
\end{align}
Plugging (\ref{N=1 Fefferman-Graham expansions}) into the bulk on-shell action, we can read off its divergent part as $\varepsilon\to 0$, which is canceled by the following counterterm,
\begin{align}
    I_{\text{ct}}=&\frac{1}{8\pi G}\int_{\partial\mathcal{M}_{\varepsilon}}d^2x\sqrt{\gamma}\Big[1+\frac{1}{M}\frac{8\pi G\phi^2}{\varepsilon^{-\frac{2}{M}}}+\sum_{m=3}^{M-1}\frac{(\pi G)^{\frac{m}{2}}\tilde D_{(\frac{2m}{M})}\phi^m}{\varepsilon^{-\frac{m}{M}}}\notag\\
&+\frac{1}{4}\Big(R_{(\gamma)}+\varepsilon(\pi G)^\frac{M}{2}\tilde{D}_{(2)}\phi^M\Big)\text{ln}\varepsilon\Big]. \label{N=1 counterterm}
\end{align}
Unlike the counterterm (\ref{N>1 counterterm}), the subleading terms here have a finite contribution to the renormalized on-shell action. More precisely, the finite part of $\phi^m\varepsilon^{\frac{m}{M}-1}$ is proportional to $\phi^M_{(0)}$ (using the FG coefficients in appendix \ref{appendix Holo Ren}). Thus, the renormalized one-point correlators take the following forms
\begin{align}
    \langle{T_{ij}}\rangle=&\frac{1}{8\pi G}\Big(g_{(2)ij}-\text{Tr}g_{(2)}g_{(0)ij}\Big)-\frac{2}{M}\phi_{(0)}\phi_{(2-\frac{4}{M})}g_{(0)ij}+(\pi G)^{\frac{M}{2}-1}\mathcal{A}_{M}\phi^M_{(0)}g_{(0)ij},\notag\\
     \langle{O}\rangle=&2(1-\frac{2}{M})\phi_{(2-\frac{4}{M})}+(\pi G)^{\frac{M}{2}-1}\mathcal{B}_{M}\phi^{M-1}_{(0)}, \label{N=1 one-point correlators}
\end{align}
where $\mathcal{A}_M$ and $\mathcal{B}_M$ are some constants. Since the bulk action (\ref{3d action}) is invariant under $\Phi\to -\Phi$, additional terms in (\ref{N=1 one-point correlators}) occur only when $M$ is even. For example, when $N=1$ and $M=4$, the counterterm is constructed as 
\begin{align}
    I_{\text{ct}}=\frac{1}{8\pi G}\int_{\partial\mathcal{M}_c}d^2x\sqrt{\gamma}\Big[1+2\pi G\varepsilon^{\frac{1}{2}}\phi^2+\frac{1}{4}R_{(\gamma)}\text{ln}\varepsilon+2\pi^2G^2\phi^4\varepsilon\text{ln}\varepsilon\Big],
\end{align}
and the renormalized one-point correlators are
\begin{align}
    \langle{T_{ij}}\rangle=&\frac{1}{8\pi G}\Big(g_{(2)ij}-\text{Tr}g_{(2)}g_{(0)ij}\Big)-\frac{1}{2}\phi_{(0)}\phi_{(1)}g_{(0)ij}+\frac{7\pi G}{4}\phi_{(0)}^4g_{(0)ij},\notag\\
    \langle{O}\rangle=&\phi_{(1)}+2\pi G\phi_{(0)}^3.
\end{align}
These terms contribute an additional trace anomaly,
\begin{align}
   \langle{T^i_{i}}\rangle=&-\frac{1}{16\pi G}R_{(0)}+\frac{2}{M}\phi_{(0)}\langle{O}\rangle+(\pi G)^{\frac{M}{2}-1}\mathcal{C}_M\phi_{(0)}^M,
\end{align}
which is also proportional to $\phi^M_{(0)}$. Moreover, one can easily prove that the conservation equation takes the same form as (\ref{massless conservation equation}). To calculate the higher-point mixed correlators, we perturbed the boundary values as (\ref{perturbed boundary values}) and expanded the one-point correlators as (\ref{perturbed one-point correlators}). The differential equations for $\langle{T_{ij}}\rangle^{[k,l]}$ are obtained from the trace relation and the conservation equation,
\begin{align}
\langle{T_{z\bar z}}\rangle^{[k,l]}=&-\frac{1}{4}\sum_{m=1}^lg^{[0,m]ij}_{(0)}\langle{T_{ij}}\rangle^{[k,l-m]}+\frac{1}{2M}\langle{O}\rangle^{[k-1,l]}f-\frac{1}{64\pi G}\delta_{k,0}R^{[0,l]}_{(0)}\notag\\
&+\delta_{k,M}\delta_{l,0}(\pi G)^{\frac{M}{2}-1}\mathcal{D}^{T_{z\bar z}}_{M}f^M,\notag\\
\partial_{\bar z}\langle{T_{zz}}\rangle^{[k,l]}=&\frac{1}{4}\sum_{m=1}^{l}[\partial_zg^{[0,m]ij}_{(0)}+g^{[0,m]ij}_{(0)}\partial_z-2\delta^i_z\nabla^{[0,m]j}_{(0)}]\langle{T_{ij}}\rangle^{[k,l-m]}+\frac{1}{2}(1-\frac{1}{M})\langle{O}\rangle^{[k-1,l]}\partial_zf\notag\\
&-\frac{1}{2M}\partial_z\langle{O}\rangle^{[k-1,l]}f+\frac{1}{64\pi G}\delta_{k,0}\partial_zR^{[0,l]}_{(0)}+\delta_{k,M}\delta_{l,0}(\pi G)^{\frac{M}{2}-1}\mathcal{D}^{T_{zz}}_{M}f^{M-1}\partial_zf,\notag\\
\partial_{z}\langle{T_{\bar z\bar z}}\rangle^{[k,l]}=&\text{ c.c. of }\partial_{\bar z}\langle{T_{zz}}\rangle^{[k,l]}. \label{N=1 higher-point mixed correlators}
\end{align}
Additional terms proportional to $\delta_{k,M}\delta_{l,0}$ in (\ref{N=1 higher-point mixed correlators}) contribute contact terms to the mixed correlators when the number of scalar operators $k\geq M$.

\section{Correlators in holographic CFT with a finite cutoff}\label{4}
In the preceding sections, we have computed the mixed correlators within the framework of AdS$_3$/CFT$_2$ and successfully derived the appropriate Ward identities. In what follows, we will be interested in the holographic aspects of the AdS$_3$ with Dirichlet boundary at finite cutoff. $(y,z,\bar z)$ are the FG coordinates in bulk, and the boundary $\partial\mathcal{M}_c$ is a hard radial cutoff at $y=y_c$. A natural holographic dictionary is given by the generalized GKPW relation,
\begin{align}
    Z_{G}[\phi_{(c)},g_{(c)ij}]=\Big\langle{\text{exp}\int_{\partial\mathcal{M}_c}d^2x\sqrt{g_{(c)}}(\phi_{(c)}O-\frac{1}{2}g^{ij}_{(c)}T_{ij})}\Big\rangle_{\text{EFT}}, \label{generalized GKPW relation}
\end{align}
where $\phi_{(c)}(z)=\phi(y_c,z)$ and $g_{(c)ij}(z)=g_{ij}(y_c,z)$. We consider the semiclassical limit and assume that the gravitational partition function is dominated by only one saddle. Then the left-hand side $Z_{G}[\phi_{(c)},g_{(c)ij}]\approx e^{-I_{\text{on-shell}}[\phi_{(c)},g_{(c)ij}]}$. The bulk action takes the same form as (\ref{3d action}) and one can introduce the counterterm on $\partial\mathcal{M}_c$,
\begin{align}
    I_{\text{ct}}=\frac{1}{8\pi G}\int_{\partial\mathcal{M}_c}d^2z\sqrt{\gamma}\Big[1+\frac{1}{4}R_{(\gamma)}\text{ln}y_c+\Psi\Big]. \label{general cutoff counterterm}
\end{align}
Here $\Psi$ is some local function of $(y_c,\phi,\partial_i\phi,g_{ij})$, whose specific form is obtained by holographic renormalization. For example, in massless scalar field coupling $\Psi=2\pi Gy_c\text{ln}y_cg^{kl}\partial_k\phi\partial_l\phi$. The renormalized one-point correlators take the forms
\begin{align}
    \langle{T_{ij}}\rangle=&\frac{1}{8\pi G}\Big[\partial_yg_{ij}-g^{kl}\partial_yg_{kl}g_{ij}\Big]_{y=y_c}-X_{ij},\notag\\
    \langle{O}\rangle=&\Big[2y^{\frac{2-\Delta}{2}}\partial_y\Big(y^{\frac{2-\Delta}{2}}\phi\Big)\Big]_{y=y_c}-Y, \label{general case renormalized one-point correlator}
\end{align}
where we have used the notation 
\begin{align}
    X_{ij}=&\frac{1}{8\pi Gy_c}\Big[\Psi g_{ij}-2\frac{\partial\Psi}{\partial g^{ij}}\Big]_{y=y_c},\notag\\
    Y=&\frac{1}{8\pi Gy_c}\Big[\frac{\partial\Psi}{\partial\phi}-\frac{1}{\sqrt{g}}\partial_k(\sqrt{g}\frac{\partial\Psi}{\partial(\partial_k\phi)})\Big]_{y=y_c}.
\end{align}
Plugging (\ref{general case renormalized one-point correlator}) into (\ref{massive Einstein's equations and scalar EOM}), we obtain the trace relation
\begin{align}
    \langle{T^i_i}\rangle=&4\pi Gy_c\Big[(\langle{T_{ij}}\rangle+X_{ij})^2-(\langle{T^i_i}\rangle+X^i_i)^2\Big]+\frac{1}{2}y_c^{\Delta-1}(\langle{O}\rangle+Y)^2\notag\\
    &-\frac{1}{2}y_c^{2-\Delta}\partial^i\phi_{(c)}\partial_i\phi_{(c)}+\frac{\Delta(2-\Delta)}{2}y_c^{1-\Delta}\phi^2_{(c)}-\frac{R_{(c)}}{16\pi G}-X_i^i,\label{general trace relation}
\end{align}
and the conservation equation
\begin{align}
\nabla^k_{(c)}\langle{T_{ki}}\rangle=\langle{O}\rangle\partial_i\phi_{(c)}.\label{general conservation equation}
\end{align}
The indices $i,j,k,l$ are raised by $g^{ij}_{(c)}$.\par
From the field theory perspective, the dual EFT in (\ref{generalized GKPW relation}) is obtained by $T^2$ deformation of the original CFT \cite{Hartman:2018tkw}. The flow equation of the EFT action is
\begin{align}
    \frac{\partial S}{\partial\lambda}=&\int d^2z\sqrt{g_{(c)}}\Big[(T_{ij}+X_{ij})^2-(T^i_i+X^i_i)^2+\frac{1}{4\pi Gy_c}\Big(t^y_y-\frac{R_{(c)}}{16\pi G}-X^i_i\Big)\Big]. \label{field theory flow equation 1}
\end{align}
Here $\lambda=2\pi Gy_c$ is the deformation parameter. $t^y_y$ is the radial component of the matter stress tensor. For example, when the coupled scalar field is massless, we have
\begin{align}
   t^y_y=&-\frac{1}{2}\partial^i\phi_{(c)}\partial_i\phi_{(c)}+\frac{y_c}{2}\Big(O-\frac{\text{ln}y_c}{2}\Box_{(c)}\phi_{(c)}\Big)^2,\notag\\
    X_{ij}=&-\frac{\text{ln}y_c}{2}\Big(\partial_i\phi_{(c)}\partial_j\phi_{(c)}-\frac{1}{2}\partial^k\phi_{(c)}\partial_k\phi_{(c)} g_{(c)ij}\Big).
\end{align}
By turning off the sources, the flow equation can be simplified as follows,
\begin{align}
    \frac{\partial S}{\partial\lambda}\Big|_{\substack{\phi_{(c)}=0\\g_{(c)ij}=\eta_{ij}}}=&\int d^2z\sqrt{\eta}\Big[T_{ij}T^{ij}-(T^i_i)^2+\frac{1}{8\pi G}O^2\Big].
\end{align}
It can be observed that, in the presence of a matter field in the bulk, the dual field theory at a finite cutoff deviates from being a standard $T\bar T$-deformed CFT. The deformation operator encompasses both the $T\bar T$ operator and the dual scalar operator.
\par
According to the generalized GKPW relation, holographic correlators are computed by (\ref{holographic correlator}), by replacing $\phi_{(0)}$ and $g_{(0)ij}$ with $\phi_{(c)}$ and $g_{(c)ij}$ respectively.
We change the boundary values like (\ref{perturbed boundary values}) and solve (\ref{general trace relation}) and (\ref{general conservation equation}) perturbatively. The terms of order $\epsilon_1^k\epsilon_2^l$ read
\begin{align}
    \langle{T_{z\bar z}}\rangle^{[k,l]}=&\bar A\langle{T_{zz}}\rangle^{[k,l]}+A\langle{T_{\bar z\bar z}}\rangle^{[k,l]}+B\langle{O}\rangle^{[k,l]}+\mathcal{F}^{[k,l]}_{T_{z\bar z}},\label{general Tzzbar kl}\\
    \partial_{\bar z}\langle{T_{zz}}\rangle^{[k,l]}=&-\partial_{z}\langle{T_{z\bar z}}\rangle^{[k,l]}+\mathcal{F}^{[k,l]}_{T_{zz}},\label{general Tzz kl}\\
    \partial_{z}\langle{T_{\bar z\bar z}}\rangle^{[k,l]}=&-\partial_{\bar z}\langle{T_{z\bar z}}\rangle^{[k,l]}+\mathcal{F}^{[k,l]}_{T_{\bar z\bar z}}.\label{general Tzbarzbar kl}
\end{align}
Here $\mathcal{F}^{[k,l]}_{T_{z\bar z}}$, $\mathcal{F}^{[k,l]}_{T_{zz}}$ and $\mathcal{F}^{[k,l]}_{T_{\bar z\bar z}}$ consist of lower-order coefficients and local functions of $f$ and $\chi_{ij}$. $A$, $\bar A$, and $B$ are constants determined by the one-point correlators at the saddle,
\begin{align}
    A=&\frac{8\pi Gy_c\langle{T_{zz}}\rangle^{[0,0]}}{1+16\pi Gy_c\langle{T_{z\bar z}}\rangle^{[0,0]}},\ \ \bar A=\frac{8\pi Gy_c\langle{T_{\bar z\bar z}}\rangle^{[0,0]}}{1+16\pi Gy_c\langle{T_{z\bar z}}\rangle^{[0,0]}},\notag\\
    B=&\frac{y_c^{\Delta-1}\langle{O}\rangle^{[0,0]}}{4(1+16\pi Gy_c\langle{T_{z\bar z}}\rangle^{[0,0]})}.
\end{align}
Plugging (\ref{general Tzzbar kl}) into (\ref{general Tzz kl}) and (\ref{general Tzbarzbar kl}) we obtain
\begin{align}
    (\partial_{\bar z}+\bar A\partial_z)\langle{T_{zz}}\rangle^{[k,l]}+A\partial_z\langle{T_{\bar z\bar z}}\rangle^{[k,l]}=&-B\partial_z\langle{O}\rangle^{[k,l]}-\partial_{z}\mathcal{F}^{[k,l]}_{T_{z\bar z}}+\mathcal{F}^{[k,l]}_{T_{zz}},\label{coupled equation 1}\\
    \bar A\partial_{\bar z}\langle{T_{zz}}\rangle^{[k,l]}+(\partial_z+A\partial_{\bar z})\langle{T_{\bar z\bar z}}\rangle^{[k,l]}=&-B\partial_{\bar z}\langle{O}\rangle^{[k,l]}-\partial_{\bar z}\mathcal{F}^{[k,l]}_{T_{z\bar z}}+\mathcal{F}^{[k,l]}_{T_{\bar z\bar z}}.\label{coupled equation 2}
\end{align}
Adding $(\partial_z+A\partial_{\bar z})$(\ref{coupled equation 1}) and $-A\partial_z$(\ref{coupled equation 2}) we find
\begin{align}
    &(\partial_z\partial_{\bar z}+\bar A\partial^2_{z}+A\partial^2_{\bar z})\langle{T_{zz}}\rangle^{[k,l]}\notag\\
    =&-B\partial_z^2\langle{O}\rangle^{[k,l]}-\partial_z^2\mathcal{F}^{[k,l]}_{T_{z\bar z}}+(\partial_z+A\partial_{\bar z})\mathcal{F}^{[k,l]}_{T_{zz}}-A\partial_z\mathcal{F}^{[k,l]}_{T_{\bar z\bar z}}. \label{decoupled equation 1}
\end{align}
The matrix $M$ defined by $M^{ij}\partial_{i}\partial_{j}=\partial_z\partial_{\bar z}+\bar A\partial^2_{z}+A\partial^2_{\bar z}$ is positive definite\footnote{$M$ is always positive definite in our consideration. Using the trace relation (\ref{general trace relation}) we find
\begin{align}
    |A|^2=\frac{1}{4+\frac{1+4\pi Gy_c^{\Delta}(\langle{O}\rangle^{[0,0]})^2}{|8\pi Gy_c\langle{T_{zz}}\rangle^{[0,0]}|^2}}<\frac{1}{4}.
\end{align}} as long as $|A|<\frac{1}{2}$. We introduce the following coordinate transformation,
\begin{align}
    z=&\frac{\bar A+\tilde{A}}{4\tilde{A}}\Big[(1+\Delta_A+\frac{i(1+\Delta_A)\text{Im}A}{\Delta_A(\text{Re}A+\tilde{A})})Z+(1-\Delta_A-\frac{i(1-\Delta_A)\text{Im}A}{\Delta_A(\text{Re}A+\tilde{A})})\bar Z\Big],\notag\\
    \bar z=&\frac{A+\tilde{A}}{4\tilde{A}}\Big[(1-\Delta_A+\frac{i(1-\Delta_A)\text{Im}A}{\Delta_A(\text{Re}A+\tilde{A})})Z+(1+\Delta_A-\frac{i(1+\Delta_A)\text{Im}A}{\Delta_A(\text{Re}A+\tilde{A})})\bar Z\Big], \label{patch 1}
\end{align}
where $\tilde{A}=\text{sgn}(\text{Re}A)|A|$ and $\Delta_A=\sqrt{\frac{1-2\tilde{A}}{1+2\tilde{A}}}$. The periods in $Z$ are $1$ and $\Omega$ with the new modular parameter
\begin{align}
    \Omega=\frac{(\Delta_A+1)(A+\tilde{A})\tau+(\Delta_A-1)(\bar A+\tilde{A})\bar\tau}{2\Delta_A(A_1+\tilde{A})+2i\text{Im}A}.
\end{align}
Plugging (\ref{patch 1}) into the left-hand side of (\ref{decoupled equation 1}) we obtain
\begin{align}
    (\partial_z\partial_{\bar z}+\bar A\partial^2_{z}+A\partial^2_{\bar z})\langle{T_{zz}}\rangle^{[k,l]}=&\frac{2\Delta_A^2\tilde A(1+2\tilde A)(\text{Re}A+\tilde A)}{\Delta_A^2(\text{Re}A+\tilde A)^2+\text{Im}A^2}\partial_Z\partial_{\bar Z}\langle{T_{zz}}\rangle^{[k,l]}.
\end{align}
Then equation (\ref{decoupled equation 1}) can be solved by the torus Green's function (which is defined in appendix \ref{appendix torus Green's function}), 
\begin{align}
    &\frac{\delta^{k+l}\langle{T_{zz}(z)}\rangle^{[k,l]}}{\prod_{i=1}^k\delta f(x_i)\prod_{j=1}^l\delta \chi_{\alpha_j\beta_j}(y_j)}\notag\\
    =&\int d^2W\Big(\frac{1}{\text{Im}\Omega}+\frac{1}{\pi}\tilde{G}_{\Omega}(W-Z)\partial_W\partial_{\bar W}\Big)\frac{\delta^{k+l}\langle{T_{zz}(w)}\rangle^{k,l}}{\prod_{i=1}^k\delta f(x_i)\prod_{j=1}^l\delta \chi_{\alpha_j\beta_j}(y_j)}. \label{general result}
\end{align}
Later, we will perform this procedure to compute the specific holographic correlators. Note that (\ref{patch 1}) is available in the domain $\lbrace{A,|A|<\frac{1}{2}}\rbrace\backslash\lbrace A,\text{Re}A=0\text{ and }0<|\text{Im}A|<\frac{1}{2} \rbrace$. When $A$ is a pure imaginary number, we can apply the following coordinate transformation,
\begin{align}
    z=&\frac{(i+\Delta'_A)^2|A|+[1+(\Delta'_A)^2]\bar A}{4i\Delta'_A|A|}Z'-\frac{(-i+\Delta'_A)^2|A|+[1+(\Delta'_A)^2]\bar A}{4i\Delta'_A|A|}\bar Z',\notag\\
    \bar z=&\frac{(i+\Delta'_A)^2|A|+[1+(\Delta'_A)^{2}]A}{4i\Delta'_A|A|}Z'-\frac{(-i+\Delta'_A)^2|A|+[1+(\Delta'_A)^2]A}{4i\Delta'_A|A|}\bar Z', \label{patch 2}
\end{align}
where $\Delta'_A=\sqrt{\frac{(1+2|A|)(|A|-\text{Re}A)}{(1-2|A|)(|A|+\text{Re}A)}}$. The periods in $Z'$ are $1$ and $\Omega'=\frac{1}{\text{Im}A}[\text{Im}(A\tau)+\frac{-i+\Delta'_A}{i+\Delta'_A}|A|\text{Im}\tau]$. The coordinate transformation (\ref{patch 2}) fails when $A$ is a real number.\par
To end this section, let us discuss the ambiguity of the counterterm, as also mentioned in \cite{Hartman:2018tkw}. It should be noted that the counterterm (\ref{general cutoff counterterm}) lacks covariance due to its explicit dependence on $y_c$. Consequently, counterterms of the same form constructed in alternative FG coordinate systems exhibit differences from (\ref{general cutoff counterterm}), which tend to zero as $y_c\to 0$. More generally, any counterterm that effectively cancels the divergence of the on-shell action in the CFT limit is permissible. Diverse choices of counterterms yield distinct holographic trace relations, resulting in varied flow equations for the boundary field theory.

\subsection{Two-point correlators on general saddles}\label{4.1}
For simplicity, we consider the massless scalar field coupling in the remainder of this paper. From the counterterm (\ref{general cutoff counterterm}) with $\Psi=2\pi Gy_c\text{ln}y_cg^{kl}\partial_k\phi\partial_l\phi$, it follows that
\begin{align}
    X_{ij}=&-\frac{\text{ln}y_c}{2}\Big(\partial_i\phi_{(c)}\partial_j\phi_{(c)}-\frac{1}{2}\partial^k\phi_{(c)}\partial_k\phi_{(c)} g_{(c)ij}\Big),\notag\\
    Y=&-\frac{\text{ln}y_c}{2}\Box_{(c)}\phi_{(c)}. \label{general cutoff massless X and Y}
\end{align}
The two-point mixed correlators on a general saddle are computed by implementing the procedure illustrated in (\ref{general Tzzbar kl})-(\ref{general result}). The results are listed as follows:
\begin{align}
    \langle{T_{zz}(z)O(z_1)}\rangle=&\frac{-i[(1-A)\partial_{\tau}-A\partial_{\bar\tau}]\langle{O}\rangle}{1-2\text{Re}A}+\frac{(\Delta_A^2+\tilde{A}-\text{Re}A)\text{Im}(A\tau)}{2\pi\Delta_A|A|^2(1-2\text{Re}A)\text{Im}\tau}\notag\\
    &\times\Big\lbrace\int d^2w\Big[(\partial_w^2\tilde{G}_{\Omega}(W-Z)-\frac{2\pi\Delta_A|A|^2}{(\Delta_A^2+\tilde{A}-\text{Re}A)\text{Im}(A\tau)})\notag\\
    &\times(\frac{\langle{O}\rangle}{2}\delta(w-z_1)-B\langle{O(w)O(z_1)}\rangle)\Big]-2B\text{ln}y_c\partial_{z_1}^2\partial_{\bar z_1}\tilde{G}_{\Omega}(Z_1-Z)\Big\rbrace,\notag\\
    \langle{T_{\bar z\bar z}(z)O(z_1)}\rangle=&\text{ c.c. of }\langle{T_{zz}(z)O(z_1)}\rangle,\notag\\
        \langle{T_{z\bar z}(z)O(z_1)}\rangle=&\bar A\langle{T_{zz}(z)O(z_1)}\rangle+A\langle{T_{\bar z\bar z}(z)O(z_1)}\rangle+B\langle{O(z)O(z_1)}\rangle\notag\\
        &-2B\text{ln}y_c\partial_z\partial_{\bar z}\delta(z-z_1)].\label{general cutoff 2pt mixed 4}
\end{align}
We also compute the two-point stress tensor correlators,
\begin{align}
&\langle{T_{zz}(z)T_{zz}(z_1)}\rangle\notag\\
    =&\frac{-i[(1-A)\partial_{\tau}-A\partial_{\bar\tau}]\langle{T_{zz}}\rangle}{1-2\text{Re}A}-\frac{(1-A)\langle{T_{zz}}\rangle}{1-2\text{Re}A}\notag\\
    &+\frac{(\Delta_A^2+\tilde{A}-\text{Re}A)\text{Im}(A\tau)}{4\pi\Delta_A|A|^2(1-2\text{Re}A)\text{Im}\tau}\Big\lbrace\frac{1}{1+16\pi Gy_c\langle{T_{z\bar z}}\rangle}\Big[2\langle{T_{zz}}\rangle\partial_{z_1}\notag\\
    &+16\pi Gy_c(2\langle{T_{z\bar z}}\rangle\partial_{z_1}+\langle{T_{zz}}\rangle\partial_{\bar{z}_1})-\frac{1}{16\pi G}\partial^3_{z_1}\Big]\partial_{z_1}\tilde{G}_{\Omega}(Z_1-Z)\notag\\
    &-2B\int d^2w\Big[(\partial_w^2\tilde{G}_{\Omega}(W-Z)-\frac{2\pi\Delta_A|A|^2}{(\Delta_A^2+\tilde{A}-\text{Re}A)\text{Im}(A\tau)})\langle{O(w)T_{zz}(z_1)}\rangle\Big]\Big\rbrace,\notag\\
    &\langle{T_{zz}(z)T_{\bar z\bar z}(z_1)}\rangle\notag\\
    =&\frac{-i[(1-A)\partial_\tau-A\partial_{\bar\tau}]\langle{T_{\bar z\bar z}}\rangle}{1-2\text{Re}A}+\frac{(1-2\text{Re}A)\langle{T_{z\bar z}}\rangle-A\langle{T_{\bar z\bar z}}\rangle}{\text{Im}\tau(1-2\text{Re}A)}\notag\\
    &+\frac{(\Delta_A^2+\tilde{A}-\text{Re}A)\text{Im}(A\tau)}{4\pi\Delta_A|A|^2(1-2\text{Re}A)\text{Im}\tau}\Big\lbrace\frac{1}{1+16\pi Gy_c\langle{T_{z\bar z}}\rangle}\Big[16\pi Gy_c\langle{T_{z\bar z}}\rangle(\langle{T_{\bar z\bar z}}\rangle\partial_{z_1}^2\notag\\
    &+\langle{T_{z\bar z}}\rangle\partial_{z_1}\partial_{\bar z_1}+\langle{T_{zz}}\rangle\partial_{\bar{z}_1}^2)+\frac{y_c}{4}\langle{O}\rangle^2\partial_{z_1}\partial_{\bar z_1}-\frac{1}{16\pi G}\partial^2_{z_1}\partial^2_{\bar z_1}\Big]\tilde{G}_{\Omega}(Z_1-Z)\notag\\
    &-2B\int d^2w\Big[(\partial_w^2\tilde{G}_{\Omega}(W-Z)-\frac{2\pi\Delta_A|A|^2}{(\Delta_A^2+\tilde{A}-\text{Re}A)\text{Im}(A\tau)})\langle{O(w)T_{\bar z\bar z}(z_1)}\rangle\Big]\Big\rbrace,\notag\\
    &\langle{T_{z\bar z}(z)T_{zz}(z_1)}\rangle\notag\\
    =&\bar A\langle{T_{zz}(z)T_{zz}(z_1)}\rangle+A\langle{T_{\bar z\bar z}(z)T_{zz}(z_1)}\rangle+B\langle{O(z)T_{zz}(z_1)}\rangle\notag\\
    &+\frac{1}{2(1+16\pi Gy_c\langle{T_{z\bar z}}\rangle)}(\langle{T_{zz}}\rangle+\frac{1}{16\pi G}\partial_{z}^2)\delta(z-z_1),\notag\\
     &\langle{T_{z\bar z}(z)T_{z\bar z}(z_1)}\rangle\notag\\
   =&\bar A\langle{T_{zz}(z)T_{z\bar z}(z_1)}\rangle+A\langle{T_{\bar z\bar z}(z)T_{z\bar z}(z_1)}\rangle+B\langle{O(z)T_{z\bar z}(z_1)}\rangle\notag\\
     &-\frac{1}{2(1+16\pi Gy_c\langle{T_{z\bar z}}\rangle)}(2\langle{T_{z\bar z}}\rangle-\frac{y_c}{4}\langle{O}\rangle^2+\frac{1}{16\pi G}\partial_z\partial_{\bar z})\delta(z-z_1),\notag\\
         &\langle{T_{\bar z\bar z}(z)T_{\bar z\bar z}(z_1)}\rangle=\text{ c.c. of }\langle{T_{zz}(z)T_{zz}(z_1)}\rangle,\notag\\
          &\langle{T_{z\bar z}(z)T_{\bar z\bar z}(z_1)}\rangle=\text{ c.c. of }\langle{T_{z\bar z}(z)T_{zz}(z_1)}. \label{general two-point stress tensor}
\end{align}
As a special case, we plug $A=-\frac{\pi^2y_c}{1+2\pi^2y_c}$ and $B=0$ into (\ref{general two-point stress tensor}) to derive the two-point stress tensor correlators in the cutoff thermal AdS$_3$ \cite{He:2023hoj}.

\subsection{Three-point mixed correlators on thermal AdS$_{3}$ saddle}\label{4.2}
Let us now specify the saddle as thermal AdS$_3$ with the scalar field turned off,
\begin{align}
    ds^2=&\frac{d\rho^2}{4\rho^2}+\frac{1}{\rho}\Big[dZd\bar Z-\pi^2\rho(dZ^2+d\bar Z^2)+\pi^4\rho^2dZd\bar Z\Big],\ \ \ \phi(\rho,Z)=0. \label{cutoff thermal AdS3}
\end{align}
The periods in $Z$ are $1$ and $\Omega$. We introduce the following coordinate transformation,
\begin{align}
    y=\frac{\rho}{(1-\pi^2\rho_c)^2},\ \ \ \ z=\frac{Z-\pi^2\rho_c\bar Z}{1-\pi^2\rho_c},\ \ \ \ \bar z=\frac{-\pi^2\rho_cZ+\bar Z}{1-\pi^2\rho_c}.\label{FG coordinate transformation}
\end{align}
The new FG coordinates $(y,z,\bar z)$ are defined in a manner that ensures the boundary metric on $\partial\mathcal{M}_c$ takes on an Euclidean form.
The periods in $z$ are $1$ and $\tau=\frac{\Omega-\pi^2\rho_c\bar\Omega}{1-\pi^2\rho_c}$.
Combining (\ref{general case renormalized one-point correlator})(\ref{general cutoff massless X and Y}) we obtain the one-point correlators
\begin{align}
        \langle{T_{zz}}\rangle=&\frac{-\pi}{8G\sqrt{1+4\pi^2y_c}},\ \ \ \langle{T_{\bar z\bar z}}\rangle=\frac{-\pi}{8G\sqrt{1+4\pi^2y_c}},\notag\\
        \langle{T_{z\bar z}}\rangle=&\frac{4\pi^3y_c}{8G\sqrt{1+4\pi^2y_c}(1+\sqrt{1+4\pi^2y_c})^2},\ \ \ \langle{O}\rangle=0. \label{finite cutoff one-point correlators 1}
\end{align}
Plugging (\ref{finite cutoff one-point correlators 1}) into (\ref{general cutoff 2pt mixed 4}) and (\ref{general two-point stress tensor}), we find that the two-point mixed correlators $\langle{T_{ij}O}\rangle$ vanish, and the stress tensor correlators $\langle{T_{ij}T_{kl}}\rangle$ consist with the results in pure gravity. We further compute the three-point mixed correlators $\langle{T_{ij}OO}\rangle$,
\begin{align}
&\langle{T_{zz}(z)O(z_1)O(z_2)}\rangle\notag\\
    =&-\frac{i}{2}\partial_{\tau}\langle{O(z_1)O(z_2)}\rangle+\partial_{z_2}\Big[(\frac{1+2\pi^2y_c}{2\pi\sqrt{1+4\pi^2y_c}}\partial_{z_2}\tilde{G}_{\Omega}(Z_2-Z)-i\frac{\text{Im}z_2}{\text{Im}\tau})\langle{O(z_1)O(z_2)}\rangle\Big]\notag\\
    &+\frac{\pi y_c}{2\sqrt{1+4\pi^2y_c}}\Big[\frac{2\pi}{\sqrt{1+4\pi^2y_c}}(\frac{i}{2}(\partial_\tau-\partial_{\bar\tau})+i\frac{\text{Im}z_2}{\text{Im}\tau}(\partial_{z_2}-\partial_{\bar z_2})+\frac{1}{\text{Im}\tau})\notag\\
    &+\partial_{z_2}\tilde{G}_{\Omega}(Z_2-Z)\partial_{\bar z_2}-\partial_{\bar z_2}\tilde{G}_{\Omega}(Z_2-Z)\partial_{z_2}\Big]\langle{O(z_1)O(z_2)}\rangle\notag\\
    &+\frac{y_c\text{ln}y_c}{2}\partial_{z_2}\partial_{\bar z_2}\Big[(\frac{1}{\pi}\partial^2_{z_2}\tilde{G}_{\Omega}(Z_2-Z)-\frac{1}{\text{Im}\tau\sqrt{1+4\pi^2y_c}})\langle{O(z_1)O(z_2)}\rangle\Big]\notag\\
    &-\frac{y_c}{8}\int d^2w\Big[(\frac{1}{\pi}\partial^2_w\tilde{G}_{\Omega}(W-Z)-\frac{1}{\text{Im}\tau\sqrt{1+4\pi^2y_c}})\langle{O(w)O(z_1)}\rangle\langle{O(w)O(z_2)}\rangle\Big]\notag\\
    &-(\partial_{\bar z_2}-\frac{\pi^2y_c\text{ln}y_c}{2\sqrt{1+4\pi^2y_c}}\partial_{z_2})\Big[(\frac{1}{\pi}\partial^2_w\tilde{G}_{\Omega}(W-Z)-\frac{1}{\text{Im}\tau\sqrt{1+4\pi^2y_c}})\partial_{z_2}\delta(z_1-z_2)\Big]\notag\\
    &+\frac{\pi^2 y_c\text{ln}y_c}{2\sqrt{1+4\pi^2}}\partial_{\bar z_2}\Big[(\frac{1}{\pi}\partial^2_w\tilde{G}_{\Omega}(W-Z)-\frac{1}{\text{Im}\tau\sqrt{1+4\pi^2y_c}})\partial_{\bar z_2}\delta(z_1-z_2)\Big]\notag\\
    &-\frac{y_c(\text{ln}y_c)^2}{2}\partial_{z_2}\partial_{\bar z_2}\Big[(\frac{1}{\pi}\partial^2_w\tilde{G}_{\Omega}(W-Z)-\frac{1}{\text{Im}\tau\sqrt{1+4\pi^2y_c}})\partial_{z_2}\partial_{\bar z_2}\delta(z_1-z_2)\Big]+(z_1\leftrightarrow z_2),\notag\\
&\langle{T_{z\bar z}(z)O(z_1)O(z_2)}\rangle\notag\\
=&-\frac{\pi^2y_c}{2(1+2\pi^2y_c)}\Big[\langle{T_{zz}(z)O(z_1)O(z_2)}\rangle+\langle{T_{\bar z\bar z}(z)O(z_1)O(z_2)}\rangle\Big]\notag\\
&+\frac{1}{2(1+2\pi^2y_c)}\Big[\sqrt{1+4\pi^2y_c}[\frac{y_c}{4}\langle{O(z)O(z_1)}\rangle\langle{O(z)O(z_2)}\rangle-\partial_z\delta(z-x_1)\partial_{\bar z}\delta(z-z_2)\notag\\
&-y_c\text{ln}y_c\langle{O(z)O(z_1)}\rangle\partial_z\partial_{\bar z}\delta(z-z_2)+y_c(\text{ln}y_c)^2\partial_z\partial_{\bar z}\delta(z-z_1)\partial_z\partial_{\bar z}\delta(z-z_2)]\notag\\
&+\pi^2y_c\text{ln}y_c[\partial_z\delta(z-z_1)\partial_z\delta(z-z_2)+\partial_{\bar z}\delta(z-z_1)\partial_{\bar z}\delta(z-z_2)]\Big]+(z_1\leftrightarrow z_2),\notag\\
&\langle{T_{\bar z\bar z}(z)O(z_1)O(z_2)}\rangle=\text{ c.c. of }\langle{T_{zz}(z)O(z_1)O(z_2)}\rangle.
\end{align}
The calculation of $\langle{O(z)O(z')}\rangle$ can be accomplished by following the procedure in section \ref{3.1.2}. The radial coordinate $x$ and the field redefinition $y^{[1]}_{m,n}$, introduced in (\ref{transformation}), continue to be employed. The boundary conditions are modified to
\begin{align}
    y^{[1]}_{m,n}(x)\Big|_{x\to x_c}&=\frac{f_{m,n}}{x_c^{\frac{n}{2}}(1-x_c)},\notag\\
    x^{\frac{n}{2}}y^{[1]}_{m,n}(x)\Big|_{x\to 0}&\ \ \ \ \text{regular}.
\end{align}
The exact solution takes the form
 \begin{gather}
y^{[1]}_{m,n}(x)=
\begin{cases}
\frac{f_{m,n}F(\alpha,\beta,\gamma,x)}{x_c^{\frac{n}{2}}(1-x_c)F(\alpha,\beta,\gamma,x_c)},&n\geq 0,\\
\frac{f_{m,n}x^{-n}F(\alpha-n,\beta-n,1-n,x)}{x_c^{-\frac{n}{2}}(1-x_c)F(\alpha-n,\beta-n,1-n,x_c)},&n<0.
\end{cases} \label{allowed solution 2}
\end{gather}
It follows that 
\begin{align}
    \Phi^{[1]}(x,\varphi,t)=&\sum^{+\infty}_{m,n=-\infty}\Big[\frac{x^{\frac{|n|}{2}}(1-x)F(\tilde\alpha,\tilde\beta,\tilde\gamma,x)}{x_c^{\frac{|n|}{2}}(1-x_c)F(\tilde\alpha,\tilde\beta,\tilde\gamma,x_c)}f_{m,n}e^{\frac{-imt}{T}}e^{in\varphi}\Big],
\end{align}
where $\tilde\alpha=1+\frac{im}{2T}+\frac{|n|}{2}$, $\tilde\beta=1-\frac{im}{2T}+\frac{|n|}{2}$ and $\tilde\gamma=1+|n|$. Now, let us go back to the FG coordinate system $(y,z,\bar z)$ via the following transformation,
\begin{align}
    x=\Big(\frac{1-4\pi^2(1+\sqrt{1+4\pi^2y_c})^{-2}y}{1+4\pi^2(1+\sqrt{1+4\pi^2y_c})^{-2}y}\Big)^2,\ \ \varphi=\pi(z+\bar z),\ \ t=-i\pi\frac{z-\bar z}{\sqrt{1+4\pi^2y_c}}.
\end{align}
The modular parameter of the cutoff torus $\tau=iT\sqrt{1+4\pi^2y_c}$. The first-order perturbed one-point correlator $\langle{O}\rangle^{[1]}$ is calculated using (\ref{general case renormalized one-point correlator}), and then the functional derivative w.r.t. $f(z')$ is taken to obtain the two-point scalar correlator,
\begin{align}
     &\langle{O(z)O(z')}\rangle\notag\\
        =&\frac{8\pi^2}{|\tau|}\sum^{+\infty}_{m,n=-\infty}\Big\lbrace\frac{1}{1+4\pi^2y_c}\Big[\frac{1}{4\pi^2y_c}-\frac{\tilde\alpha\tilde\beta F(\tilde\alpha+1,\tilde\beta+1,\tilde\gamma+1,\frac{1}{1+4\pi^2y_c})}{\tilde\gamma (1+4\pi^2y_c)F(\tilde\alpha,\tilde\beta,\tilde\gamma,\frac{1}{1+4\pi^2y_c})}-\frac{|n|}{2}\Big]\notag\\
      &-\frac{1}{\sqrt{1+4\pi^2y_c}}(\frac{n^2}{4}+\frac{m^2}{4|\tau|^2})\text{ln}y_c\Big\rbrace e^{i\pi(n-\frac{m}{\tau})(z-z')}e^{i\pi(n+\frac{m}{\tau})(\bar z-\bar z')}.\label{scalar two-point correlator at finite cutoff}
\end{align}\par
To end this section, let us discuss the non-locality of holographic correlators. We consider the thermal AdS$_3$ saddle, whose scalar one-point correlator $\langle{O}\rangle=0$ and scalar two-point correlator is explicitly obtained in (\ref{scalar two-point correlator at finite cutoff}). As previously mentioned, the two-point mixed correlators $\langle{T_{ij}O}\rangle=0$, and the stress tensor correlators $\langle{T_{ij}T_{kl}}\rangle$ are consistent with the results obtained in the absence of a matter field \cite{He:2023hoj}. Hence, no indication of non-locality can be discerned in these two-point correlators. Non-locality may potentially become evident in the higher-order correlators. For example, the three-point mixed correlator $\langle{T_{zz}(z)O(z_1)O(z_2)}\rangle$ involves an integral of the form
\begin{align}
    -\frac{y_c}{8}\int d^2w(\frac{1}{\pi}\partial^2_w\tilde{G}_{\Omega}(W-Z)-\frac{1}{\text{Im}\tau\sqrt{1+4\pi^2y_c}})\langle{O(w)O(z_1)}\rangle\langle{O(w)O(z_2)}\rangle, \label{nonlocal 1}
\end{align}
which could suggest the non-locality of the correlator. Similar integrals can also be observed in the three-point stress tensor correlators. One can verify that, due to the vanishing of $\langle{O}\rangle$, the three-point stress tensor correlators are consistent with the results in pure gravity. For example, the three-point correlator $\langle{T_{zz}(z)T_{zz}(z_1)T_{zz}(z_2)}\rangle$ in a cutoff thermal AdS$_3$ is given by equation (C.22) in the previous work \cite{He:2023hoj}, which contains an integral
\begin{align}
\pi G y_c\int d^2w&(\frac{1}{\pi}\partial^2_w\tilde{G}_{\Omega}(W-Z)-\frac{1}{\text{Im}\tau\sqrt{1+4\pi^2y_c}})\Big[\langle{T_{w\bar w}(w)T_{zz}(z_1)}\rangle\langle{T_{w\bar w}(w)T_{zz}(z_2)}\rangle\notag\\
&-\langle{T_{ww}(w)T_{zz}(z_1)}\rangle\langle{T_{\bar w\bar w}(w)T_{zz}(z_2)}\rangle+(z_1\leftrightarrow z_2)\Big]. \label{nonlocal 2}
\end{align}
It is noteworthy to mention that the techniques utilized in this section to compute the thermal AdS$_3$ correlators can also be applied to other gravitational saddles, such as the Euclidean BTZ black hole \cite{Banados:1992wn,Banados:1992gq}. Analogous terms as presented in (\ref{nonlocal 1}) and (\ref{nonlocal 2}) can also be found in the three-point correlators of these saddles.

\section{Conclusions and perspectives}\label{6}
In this paper, we investigate the holographic torus correlators of the scalar operators both at conformal infinity and at a finite cutoff. The three-dimensional bulk in our study contains Einstein gravity and a free scalar field. Firstly, we employ the near-boundary analysis to solve Einstein's equation and calculate the mixed correlators in two cases: when the coupled scalar field is massless and massive. Additionally, we derive recurrence relations for a specific class of higher-point correlators. Our results are consistent with the Ward identity in torus CFT, providing concrete validation of AdS$_3$/CFT$_2$ with non-trivial topology. Secondly, to fully determine the holographic correlators, we explore the global solution of the massless scalar EOM and accurately compute the scalar two-point correlator on the thermal AdS$_3$ saddle. However, determining the higher-point scalar correlators remains unresolved. A robust algorithm for computing scalar correlators on a general saddle is still lacking.\par
Utilizing our results to compute the precise correlators on a non-trivial saddle would be interesting. A well-known AdS-sliced domain wall solution is the Janus solution \cite{Bak:2003jk,Freedman:2003ax}, and its holographic correlators are calculated in \cite{Papadimitriou:2004rz,Chiodaroli:2016jod}. The authors of \cite{Bak:2007jm,Auzzi:2021nrj} investigate the AdS$_3$ Janus solution. We wonder whether there are analogous solutions in the bulk with a torus boundary. Another important direction is to introduce the scalar potential in the bulk. Exact bulk solutions in the presence of an exponential potential have been explored in numerous works \cite{Poletti:1994ff, Chan:1994qa, Chan:1995wj, Charmousis:2001nq,Charmousis:2009xr,Anabalon:2023efw}. Extending our study to the model that includes a general self-interacting potential of the scalar field is necessary.


\section*{Acknowledgments}
We want to thank Yi Li, Chen-Te Ma, Juntao Wang, and Long Zhao for their valuable discussions and comments. We are also grateful to all the organizers and participants of the "Quantum Information, Quantum Matter and Quantum Gravity" workshop (YITP-T-23-01) held at YITP, Kyoto University, where a part of this work was done.
This work is partly supported by the National Natural Science Foundation of China under Grant No.~12075101 and No.~12235016. S.H. is grateful for financial support from the Fundamental Research Funds for the Central Universities and the Max Planck Partner Group.

\appendix
\section{Green's functions on torus}\label{appendix torus Green's function}
In this appendix, we give the definitions of Green's functions $G_{\tau}(z-w)$ and $\tilde G_{\tau}(z-w)$ on a torus with modular parameter $\tau$. They satisfy the following differential equations,
\begin{align}
    \frac{1}{\pi}\partial_{\bar z}G_{\tau}(z-w)=&\delta(z-w)-\frac{1}{\text{Im}\tau},\notag\\
   \frac{1}{\pi}\partial_{z}\partial_{\bar z}\tilde G_{\tau}(z-w)=&\delta(z-w)-\frac{1}{\text{Im}\tau},
\end{align}
where $\delta(z-w)$ is the delta function with respect to the measure $d^2z=\frac{i}{2}dz\wedge{d\bar{z}}$. $G_{\tau}(z-w)$ can be represented by the Weierstrass $\zeta$-function,
\begin{align}
    G_\tau(z-w)=&\zeta_\tau(z-w)-2\zeta_\tau(\frac{1}{2})(z-w)+\frac{2\pi i}{\text{Im}\tau}\text{Im}(z-w).\label{torus Green's Function G}
\end{align}
$\zeta_\tau(z)$ is defined by a series expansion \cite{akhiezer1990elements}
\begin{align}
    \zeta_\tau(z) = \frac{1}{z} + &\sum_{(m,n)\in \mathbb{Z}^2 \setminus (0,0)} \Big(\frac{1}{z-(m+n\tau)} + \frac{1}{(m+n\tau)} + \frac{z}{(m+n\tau)^2}\Big).
\end{align}
The inclusion of the last two terms in (\ref{torus Green's Function G}) guarantees the double periodicity of $G_\tau(z-w)$. $\tilde G_{\tau}(z-w)$ can be represented by the Weierstrass $\sigma$-function,
\begin{align}
          \tilde{G}_\tau(z-w) = \text{ln}(|\sigma_\tau(z-w)|^2) - \zeta_\tau(\frac{1}{2}) (z-w)^2 - \overline{\zeta_\tau(\frac{1}{2})} (\bar{z}-\bar{w})^2 - \frac{2 \pi}{\text{Im}\tau} [\text{Im}(z-w)]^2.
\end{align}
$\sigma_\tau(z)$ is defined as
\begin{align}
    \sigma_\tau(z) = z \prod_{(m,n)\in \mathbb{Z}^2\setminus(0,0)} \Big(1-\frac{z}{m+n\tau}\Big) e^{\frac{z}{m+n\tau} + \frac{z^2}{2(m+n\tau)^2}}.
\end{align}
\section{Green's function $\mathscr{G}_{m,n}(x,x_0)$}\label{appendix Green's function 2}
In this section, we attempt to construct the exact Green's function for the hypergeometric equation, which is used to calculate the higher-order scalar coefficients in section \ref{3.1.2}. The Green's function $\mathscr{G}_{m,n}(x,x_0)$ satisfies the following differential equation
\begin{align}
    \Big(x(1-x)\partial_x^2+[\gamma-(\alpha+\beta+1)x]\partial_x-\alpha\beta\Big)\mathscr{G}_{m,n}(x,x_0)=\delta(x-x_0), \label{Green's function equation}
\end{align}
where the parameters $\alpha=1+\frac{im}{2T}+\frac{n}{2}$, $\beta=1-\frac{im}{2T}+\frac{n}{2}$ and $\gamma=1+n$ for $m,n\in\mathbb{Z}$. 
A particular solution to this differential equation can be written as
\begin{align}
    y_{m,n}^*(x,x_0)=&\frac{y^{(3)}_{m,n}(x_0)y^{(4)}_{m,n}(x_0)}{\Delta(x_0)x_0(1-x_0)}H(x-x_0)\Big(\frac{y^{(4)}_{m,n}(x)}{y^{(4)}_{m,n}(x_0)}-\frac{y^{(3)}_{m,n}(x)}{y^{(3)}_{m,n}(x_0)}\Big),
\end{align}
where $\Delta=y^{(3)}_{m,n}(y^{(4)}_{m,n})'-y^{(4)}_{m,n}(y^{(3)}_{m,n})'$ is the Wronskian. $H(x-x_0)$ is the Heaviside step function defined as $H(x-x_0)=1$ for $x>x_0$ and $H(x-x_0)=0$ for $x\leq x_0$. $y^{(3)}_{m,n}$ and $y^{(4)}_{m,n}$ are two independent solutions of the hypergeometric equation in the neighborhood of $x=1$. Note that $\gamma-\alpha-\beta\equiv-1$, these two solutions take the forms
\begin{align}
    y^{(3)}_{m,n}(x)=&F(\alpha,\beta,2,1-x),\notag\\
    y^{(4)}_{m,n}(x)=&F(\alpha,\beta,2,1-x)\text{ln}(1-x)+\frac{1}{(\alpha-1)(\beta-1)(1-x)}+\sum_{k=0}^{\infty}\frac{(\alpha)_k(\beta)_k}{k!(2)_k}\notag\\
    &\times[\psi(\alpha+k)+\psi(\beta+k)-\psi(2+k)-\psi(1+k)](1-x)^k, \label{Solutions near x=1 1}
\end{align}
where $F(\alpha,\beta,2,1-x)$ is the hypergeometric function, $(\lambda)_k=1$ for $k=0$ and $(\lambda)_k=\lambda(\lambda+1)\cdots(\lambda+k-1)$ for $k\geq 1$, and $\psi(\lambda)=\frac{\Gamma'(\lambda)}{\Gamma(\lambda)}$. If $\alpha$ and $\beta$ are negative integers, which means that $m=0$ and $n=-4,-6,-8,\cdots$, then $\psi(1+\frac{n}{2}+k)$ diverges for $k\leq -1-\frac{n}{2}$. The second solution should be modified to
\begin{align}
    y^{(4)}_{0,n}(x)=&F(1+\frac{n}{2},1+\frac{n}{2},2,1-x)\text{ln}(1-x)+\frac{4}{n^2(1-x)}+\sum_{k=0}^{-1-\frac{n}{2}}\frac{[(-1-\frac{n}{2})_k]^2}{k!(2)_k}\notag\\
    &\times[2\psi(-k-\frac{n}{2})-\psi(2+k)-\psi(1+k)](1-x)^k. \label{Solutions near x=1 2}
\end{align}
Obviously, the particular solution $y^*_{m,n}$ satisfies
\begin{align}
    (1-x)y^*_{m,n}(x,x_0)\Big|_{x\to 1}=&\frac{F(\alpha,\beta,2,1-x_0)}{\Delta(x_0)x_0(1-x_0)(\alpha-1)(\beta-1)},\notag\\
    x^{\frac{n}{2}}y^*_{m,n}(x,x_0)\Big|_{x\to 0}=&0. \label{boundary condition of particular solution}
\end{align}
Let $y^{(1)}_{m,n}$ and $y^{(2)}_{m,n}$ be two independent solutions of the hypergeometric equation in the neighborhood of $x=0$. The general solution to (\ref{Green's function equation}) takes the form
\begin{align}
    \mathscr{G}_{m,n}(x,x_0)=&A_{m,n}y^{(1)}_{m,n}(x)+B_{m,n}y^{(2)}_{m,n}(x)+y^*_{m,n}(x,x_0), \label{general form of Green's function}
\end{align}
where $A_{m,n}$ and $B_{m,n}$ are undetermined constants. As we discussed above, the Green's function $\mathscr{G}_{m,n}$ satisfies regularity condition at $x=0$
\begin{align}
    x^{\frac{n}{2}}\mathscr{G}_{m,n}(x,x_0)\Big|_{x\to 0}\ \ \ \ \text{regular}. \label{GRC for Green's function}
\end{align}
The asymptotic behaviors of the two solutions are $y^{(1)}_{m,n}\sim 1$, $y^{(2)}_{m,n}\sim x^{-n}$ for $n\geq 1$, $y^{(1)}_{m,n}\sim 1$, $y^{(2)}_{m,n}\sim \text{ln}x$ for $n=0$, and $y^{(1)}_{m,n}\sim x^{-n}$, $y^{(2)}_{m,n}\sim 1$ for $n\leq -1$. Thus the regularity condition (\ref{GRC for Green's function}) requires that $B_{m,n}=0$. $A_{m,n}$ can be determined by the boundary condition of $\mathscr{G}_{m,n}$ at $x=1$
\begin{align}
    (1-x)\mathscr{G}_{m,n}(x,x_0)\Big|_{x\to 1}=0, \label{BC of Green's function}
\end{align}
which is discussed as follows:
\begin{itemize}
\item For $n\geq 0$, the homogeneous solution $y^{(1)}_{m,n}$ takes the form 
\begin{align}
y^{(1)}_{m,n}(x)=&F(\alpha,\beta,\gamma,x).
\end{align}
The hypergeometric function can be expanded in the neighborhood of $x=1$
\begin{align}
    F(\alpha,\beta,\gamma,x)=&\frac{\Gamma(\gamma)(1-x)^{-1}}{\Gamma(\alpha)\Gamma(\beta)}+\frac{\Gamma(\gamma)\text{ln}(1-x)}{\Gamma(\alpha-1)\Gamma(\beta-1)}+O(1), \label{F's expansion in the neighborhood of x=1}
\end{align}
where $O(1)$ contains the regular terms as $x\to 1$. Plugging (\ref{boundary condition of particular solution})(\ref{general form of Green's function})(\ref{F's expansion in the neighborhood of x=1}) into (\ref{BC of Green's function}) we obtain
\begin{align}
    A_{m,n}=&-\frac{\Gamma(\alpha)\Gamma(\beta)F(\alpha,\beta,2,1-x_0)}{\Delta(x_0)x_0(1-x_0)(\alpha-1)(\beta-1)\Gamma(\gamma)}.
\end{align}
\item For $n<0$, the homogeneous solution $y_{m,n}^{(1)}$ takes the form
\begin{align}
    y_{m,n}^{(1)}(x)=&x^{-n}F(\alpha-n,\beta-n,1-n,x).
\end{align}
The expansion of $y_{m,n}^{(1)}$ in the neighborhood of $x=1$ is
\begin{align}
    y_{m,n}^{(1)}(x)=&\frac{\Gamma(1-n)(1-x)^{-1}}{\Gamma(\alpha-n)\Gamma(\beta-n)}+\frac{\Gamma(1-n)\text{ln}(1-x)}{\Gamma(\alpha-n-1)\Gamma(\beta-n-1)}+O(1). \label{y's expansion in the neighborhood of x=1}
\end{align}
Plugging (\ref{boundary condition of particular solution})(\ref{general form of Green's function})(\ref{y's expansion in the neighborhood of x=1}) into (\ref{BC of Green's function}) we obtain
\begin{align}
    A_{m,n}=&-\frac{\Gamma(\alpha-n)\Gamma(\beta-n)F(\alpha,\beta,2,1-x_0)}{\Delta(x_0)x_0(1-x_0)(\alpha-1)(\beta-1)\Gamma(1-n)}.
\end{align}
\end{itemize}
Finally, we can put everything together to obtain the explicit form of $\mathscr{G}_{m,n}(x,x_0)$. Note that (\ref{boundary condition of particular solution}) does not apply to the zero mode $y^*_{0,0}$, which satisfies $\lim_{x\to 1}(1-x)y^*_{0,0}(x,x_0)=0$. The zero mode of Green's function can be written as
\begin{align}
    \mathscr{G}_{0,0}(x,x_0)=&-H(x-x_0)\frac{(1-x_0)\text{ln}x\text{ln}(\frac{1-x}{1-x_0})}{x_0\text{ln}x_0(1-x)}.
\end{align}
\section{Review of holographic renormalization}\label{2.1}
To obtain the finite holographic correlators of boundary CFT, one needs to introduce additional boundary terms to cancel the divergence in the on-shell action. In what follows we will review this prescription in the context of AdS$_3$ coupled to a massless scalar field. The first step is to determine the coefficients in the FG expansions (\ref{massless FG expansion}). In the FG coordinate system (\ref{Fefferman-Graham coordinates}), Einstein's equations and scalar EOM can be rewritten as \cite{Henningson:1998gx,deHaro:2000vlm}
\begin{align}
&\rho(2\partial_\rho^2g_{ij}-2\partial_\rho g_{ik}g^{kl}\partial_\rho g_{lj}+g^{kl}\partial_\rho g_{kl}\partial_\rho g_{ij})+R_{(g)ij}-g^{kl}\partial_\rho g_{kl}g_{ij}+8\pi G\partial_i\phi\partial_j\phi=0,\notag\\
&\nabla_{(g)i}(g^{kl}\partial_\rho g_{kl})-\nabla_{(g)}^k(\partial_\rho g_{ki})+16\pi G\partial_\rho\phi\partial_i\phi=0,\notag\\
&g^{kl}\partial^2_\rho g_{kl}-\frac{1}{2}g^{ij}\partial_\rho g_{jk}g^{kl}\partial_\rho g_{li}+16\pi G\partial_\rho\phi\partial_\rho\phi=0,\notag\\
&\Box_{(g)}\phi+2\rho\partial_{\rho}\text{ln}g\partial_{\rho}\phi+4\rho\partial_{\rho}^2\phi=0,\label{massless Einstein's equations and scalar EOM}
\end{align}
where we have used $m=0$ and $\phi(\rho,x)=\Phi(\rho,x)$. $R_{(g)ij}$ and $\nabla_{(g)i}$ indicate the Ricci tensor and the covariant derivative operator of the metric $g_{ij}$ respectively. Plugging (\ref{massless FG expansion}) into (\ref{massless Einstein's equations and scalar EOM}) we obtain
\begin{align}
    h_{(2)ij}=&-4\pi G(\partial_i\phi_{(0)}\partial_j\phi_{(0)}-\frac{1}{2}g_{(0)}^{kl}\partial_k\phi_{(0)}\partial_l\phi_{(0)}g_{(0)ij}),\notag\\
    \psi_{(2)}=&-\frac{1}{4}\Box_{(0)}\phi_{(0)},\notag\\
    g^{ij}_{(0)}g_{(2)ij}=&\frac{1}{2}R_{(0)}+4\pi Gg_{(0)}^{kl}\partial_k\phi_{(0)}\partial_l\phi_{(0)},\notag\\
    \nabla^{j}_{(0)}g_{(2)ij}=&\partial_{i}(g_{(0)}^{kl}g_{(2)kl})+16\pi G\phi_{(2)}\partial_{i}\phi_{(0)}. \label{massless FG coefficients}
\end{align}
Here $R_{(0)}$ represents the Ricci scalar of the boundary metric $g_{(0)ij}$. The second step is to extract the divergent part of the on-shell action. We follow the regularization procedure in \cite{Henningson:1998gx}. The bulk integral is restricted to the domain $\rho\in [\varepsilon,\pi^{-2}]$, and the boundary integral is evaluated at the finite cutoff $\rho=\varepsilon$. Plugging (\ref{Einstein's equations 0})(\ref{scalar EOM}) into (\ref{3d action}) and using the FG coordinate system (\ref{Fefferman-Graham coordinates}), we obtain the regularized on-shell action
\begin{align}
    I_{\text{reg}}=&\frac{1}{8\pi G}\int_{\mathcal{M}_{\varepsilon}}d^2xd\rho\frac{1}{\rho^2}\sqrt{g}+\frac{1}{4\pi G}\int_{\partial\mathcal{M}_{\varepsilon}}d^2x\Big[\partial_{\varepsilon}\sqrt{g}-\frac{1}{\varepsilon}\sqrt{g}\Big]. \label{massless on-shell action 0}
\end{align}
Plugging (\ref{massless FG expansion})(\ref{massless FG coefficients}) into (\ref{massless on-shell action 0}) we have 
\begin{align}
    I_{\text{reg}}=&\frac{-1}{8\pi G}\int_{\partial\mathcal{M}_{\varepsilon}}d^2x\sqrt{g_{(0)}}\Big[\frac{1}{\varepsilon}+\Big(\frac{1}{4}R_{(0)}+2\pi Gg_{(0)}^{kl}\partial_k\phi_{(0)}\partial_l\phi_{(0)}\Big)\text{ln}\varepsilon\Big]+O(1). \label{massless on-shell action 1}
\end{align}
The last step is to construct the renormalized on-shell action. The counterterm $I_{ct}$ is to cancel the divergent part of (\ref{massless on-shell action 1}),
\begin{align}
        I_{\text{ct}}=&\frac{1}{8\pi G}\int_{\partial\mathcal{M}_{\varepsilon}}d^2x\sqrt{g_{(0)}}\Big[\frac{1}{\varepsilon}+\Big(\frac{1}{4}R_{(0)}+2\pi Gg_{(0)}^{kl}\partial_k\phi_{(0)}\partial_l\phi_{(0)}\Big)\text{ln}\varepsilon\Big]+\text{regular part}.\label{massless counterterm 0}
\end{align}
The selection of counterterm is not unique, since the regular parts of two allowed counterterms can be different. The authors of \cite{deHaro:2000vlm} (see also \cite{Bianchi:2001kw}) provide a method to obtain an explicit counterterm. This method requires us to invert the FG expansions (\ref{massless FG expansion}). Plugging the inverted expansions and the coefficients (\ref{massless FG coefficients}) into (\ref{massless counterterm 0}), we can rewrite $g_{(0)ij}$, $R_{(0)}$ and $\phi_{(0)}$ in terms of $g_{ij}$, $R_{(g)}$ and $\phi$. We only keep the terms that diverge as $\varepsilon\to 0$, and finally, the counterterm takes the form
\begin{align}
    I_{\text{ct}}=&\frac{1}{8\pi G}\int_{\partial\mathcal{M}_{\varepsilon}}d^2x\sqrt{\gamma}\Big[1+\Big(\frac{1}{4}R_{(\gamma)}+2\pi G\gamma^{kl}\partial_k\phi\partial_l\phi\Big)\text{ln}\varepsilon\Big], \label{massless counterterm 1}
\end{align}
where $\gamma_{ij}(\varepsilon,x)=\frac{1}{\varepsilon}g_{ij}(\varepsilon,x)$ is the induced metric on the cutoff surface $\partial\mathcal{M}_{\varepsilon}$. The renormalized on-shell action $I_{\text{ren}}=I_{\text{reg}}+I_{\text{ct}}$, which is finite as $\varepsilon\to 0$.\par
The holographic one-point correlators are defined as the functional derivatives of renormalized on-shell action,
\begin{align}
    \langle{T_{ij}}\rangle=&\lim_{\varepsilon\to 0}\Big(\frac{2}{\sqrt{\gamma}}\frac{\delta}{\delta\gamma^{ij}}(I_{\text{reg}}+I_{\text{ct}})\Big)\notag\\
    =&\frac{1}{8\pi G}\Big(g_{(2)ij}-4\pi G\partial_i\phi_{(0)}\partial_j\phi_{(0)}-2\pi G \partial^k\phi_{(0)}\partial_k\phi_{(0)}g_{(0)ij}-\frac{1}{2}R_{(0)}g_{(0)ij}\Big),\notag\\
    \langle{O}\rangle=&\lim_{\varepsilon\to 0}\Big(\frac{-1}{\varepsilon\sqrt{\gamma}}\frac{\delta}{\delta\phi}(I_{\text{reg}}+I_{\text{ct}})\Big)\notag\\
    =&2\Big(\phi_{(2)}-\frac{1}{4}\Box_{(0)}\phi_{(0)}\Big).\label{massless one-point correlators}
\end{align}
The indices $i,j,k,l$ are raised by $g^{ij}_{(0)}$. In (\ref{massless one-point correlators}), we use the fact that $\int_{\partial\mathcal{M}_{\varepsilon}}d^2x\sqrt{g}R_{(g)}$ is a topological invariant. Plugging (\ref{massless one-point correlators}) into (\ref{massless FG coefficients}), we obtain the trace relation (\ref{massless trace relation}) and the conservation equation (\ref{massless conservation equation}).
\par
In the preceding discussion, we assumed the scalar field to be massless. However, for the massive bulk scalar, the FG expansion may differ from the form given in Eq. (\ref{massless FG expansion}). Nevertheless, if an exact FG expansion exists, the counterterm can still be constructed using the procedure outlined above. In appendix \ref{appendix Holo Ren}, we will demonstrate holographic renormalization for a general rational number $\Delta$.
\section{Holographic renormalization for massive scalar field coupling}\label{appendix Holo Ren}
In this appendix, we will construct the counterterm in the case of gravity coupled to a massive scalar field. To begin with, let us assume that $N\geq 2$ (where $\frac{2-\Delta}{2}=\frac{N}{M}$ for a rational scaling dimension $\Delta$). The FG expansions of the bulk fields are shown in (\ref{N>1 Fefferman-Graham expansions}). We first compute the coefficients $g_{(2m\frac{N}{M})ij}$ and $\phi_{(2m\frac{N}{M})}$ for $1\leq m\leq\lfloor{\frac{M}{N}}\rfloor$. Plugging (\ref{N>1 Fefferman-Graham expansions}) into (\ref{massive Einstein's equations and scalar EOM}) and extracting the coefficients of order $\rho^{m\frac{N}{M}-1}$, we obtain
\begin{align}
    &2m\frac{N}{M}(m\frac{N}{M}-1)g_{(2m\frac{N}{M})ij}+F_{(2m\frac{N}{M})ij}+\frac{1}{2(m\frac{N}{M}-2)}g^{kl}_{(0)}F_{(2m\frac{N}{M})kl}g_{(0)ij}=0,\notag\\
    &2m\frac{N}{M}[(m+2)\frac{N}{M}-1]\phi_{(2m\frac{N}{M})}+m\frac{N^2}{M^2}\text{ln}g_{(2m\frac{N}{M})}\phi_{(0)}+G_{(2m\frac{N}{M})}=0, \label{little corfficients 1}
\end{align}
where $F_{(2m\frac{N}{M})ij}$ and $G_{(2m\frac{N}{M})}$ can be expressed by the lower-order coefficients,
\begin{align}
    F_{(2m\frac{N}{M})ij}=&\sum_{\substack{m_1+m_2+m_3=m\\m_i\in\lbrace{0,1,\cdots,m-1}\rbrace}}\Big[m_1m_3\frac{N^2}{M^2}\Big(g_{(2m_1\frac{N}{M})kl}g^{kl}_{(2m_2\frac{N}{M})}g_{(2m_3\frac{N}{M})ij}\notag\\
    &-2g_{(2m_1\frac{N}{M})ik}g^{kl}_{(2m_2\frac{N}{M})}g_{(2m_3\frac{N}{M})lj}\Big)-m_2\frac{N}{M}g_{(2m_1\frac{N}{M})kl}g^{kl}_{(2m_2\frac{N}{M})}g_{(2m_3\frac{N}{M})ij}\Big]\notag\\
    &-32\pi G\frac{N}{M}(1-\frac{N}{M})\sum_{\substack{m_1+m_2+m_3=m-2\\m_i\in\lbrace{0,1,\cdots,m-2}\rbrace}}\phi_{(2m_1\frac{N}{M})}\phi_{(2m_2\frac{N}{M})}g_{(2m_3\frac{N}{M})ij},\notag\\
    G_{(2m\frac{N}{M})}=&\sum_{\substack{m_1+m_2=m\\m_i\in\lbrace{1,2,\cdots,m-1}\rbrace}}m_1(1+m_2)\frac{N^2}{M^2}\text{ln}g_{(2m_1\frac{N}{M})}\phi_{(2m_2\frac{N}{M})}. \label{little coefficients 2}
\end{align}
For $m=1$ we have $g_{(2\frac{N}{M})ij}=0$ and $\phi_{(2\frac{N}{M})}=0$. For $m=2$ we have
\begin{align}
g_{(4\frac{N}{M})ij}=&-4\pi G\phi_{(0)}^2g_{(0)ij},\notag\\
\phi_{(4\frac{N}{M})}=&\frac{4\pi G}{4-\frac{M}{N}}\phi_{(0)}^3.
\end{align}
By applying (\ref{little corfficients 1}) and (\ref{little coefficients 2}) multiple times, we find that for $3\leq m\leq\lfloor{\frac{M}{N}}\rfloor$, the coefficients of the metric and the scalar field take the forms
\begin{align}
g_{(2m\frac{N}{M})ij}=&(\pi G)^{\frac{m}{2}}A_{(2m\frac{N}{M})}\phi_{(0)}^{m}g_{(0)ij},\label{little metric coefficients}\\
\phi_{(2m\frac{N}{M})}=&(\pi G)^{\frac{m}{2}}B_{(2m\frac{N}{M})}\phi_{(0)}^{m+1},\label{little scalar coefficients}
\end{align}
where $A_{(2m\frac{N}{M})}$ and $B_{(2m\frac{N}{M})}$ are some constants. Since we consider the gravity coupled to a free scalar field, the bulk action (\ref{3d action}) is invariant under $\Phi\to-\Phi$, and thus $A_{(2m\frac{N}{M})}=B_{(2m\frac{N}{M})}=0$ for $m$ is an odd number \cite{Hung:2011ta}. Plugging (\ref{N>1 Fefferman-Graham expansions}) into (\ref{massive Einstein's equations and scalar EOM}) and extracting the coefficients of order $\text{ln}\rho$, we obtain
\begin{align}
    g^{ij}_{(0)}h_{(2)ij}=0,\ \ \ \ \ \ \psi_{(2)}=0.
\end{align}
Extracting the coefficients of order $\rho^0$, we obtain
\begin{align}
     h_{(2)ij}=&0,\notag\\
    g_{(0)}^{ij}g_{(2)ij}=&\frac{1}{2}R_{(0)}-64\pi G\frac{N}{M}(1-\frac{N}{M})\phi_{(0)}\phi_{(2-4\frac{N}{M})},\notag\\
    \nabla_{(0)}^{k}g_{(2)ki}=&\partial_{i}(g^{kl}_{(0)}g_{(2)kl})+16\pi G[\frac{N}{M}\phi_{(0)}\partial_i\phi_{(2-4\frac{N}{M})}+(1-\frac{N}{M})\phi_{(2-4\frac{N}{M})}\partial_i\phi_{(0)}].
\end{align}
Following the spirit of \cite{deHaro:2000vlm}, we plug the FG expansions into the bulk on-shell action, and evaluate the boundary integral at a cutoff surface $\partial\mathcal{M}_{\varepsilon}$,
\begin{align}
I_{\text{reg}}=&\frac{1}{8\pi G}\int_{\mathcal{M}_{\varepsilon}}d^2xd\rho\sqrt{g}\Big[\frac{1}{\rho^2}+2\frac{N}{M}(1-\frac{N}{M})\frac{8\pi G\phi^2}{\rho^{2-2\frac{N}{M}}}\Big]+\frac{1}{16\pi G}\int_{\partial\mathcal{M}_{\varepsilon}}d^2x\frac{4}{\rho}\Big[\rho\partial_{\rho}\sqrt{g}-\sqrt{g}\Big]\notag\\
=&-\frac{1}{8\pi G}\int_{\partial\mathcal{M}_{\varepsilon}}d^2x\sqrt{g_{(0)}}\Big[\frac{1}{\varepsilon}-\frac{1}{2}(1-2\frac{N}{M})\frac{8\pi G\phi_{(0)}^2}{\varepsilon^{1-2\frac{N}{M}}}+\sum_{m=3}^{\lfloor{\frac{M}{N}}\rfloor}\frac{(\pi G)^{\frac{m}{2}}C_{(2m\frac{N}{M})}\phi^{m}_{(0)}}{\varepsilon^{1-m\frac{N}{M}}}\notag\\
&+\frac{1}{4}R_{(0)}\text{ln}\varepsilon\Big]+O(1). \label{massive regularized on-shell action}
\end{align}
Plugging the inverted FG expansions into (\ref{massive regularized on-shell action}) and keeping the terms that diverge as $\varepsilon\to 0$, we obtain the allowed counterterm (\ref{N>1 counterterm}) for $N\geq 2$.\par
For the case $N=1$, the two homogeneous solutions of the scalar field overlap on the order $\rho^{1-\frac{2}{M}}$. Plugging the FG expansions (\ref{N=1 Fefferman-Graham expansions}) into (\ref{massive Einstein's equations and scalar EOM}), we find that (\ref{little metric coefficients}) holds for $2\leq m\leq M-1$, and (\ref{little scalar coefficients}) holds for $1\leq m\leq M-3$. The order $\rho^{-\frac{2}{M}}$ of the scalar EOM gives the scalar conformal anomaly,
\begin{align}
    \psi_{(2-\frac{4}{M})}=&-\frac{M}{2(M-2)}\Big[\frac{M-2}{M^2}\text{ln}g_{(2-\frac{4}{M})}\phi_{(0)}+G_{(2-\frac{4}{M})}\Big]\notag\\
    \propto&\ (\pi G)^{\frac{M-2}{2}}\phi^{M-1}_{(0)}. \label{N=1 scalar conformal anomaly}
\end{align}
The order $\text{ln}\rho$ of Einstein's equations gives the trace part of the metric conformal anomaly,
\begin{align}
    g^{ij}_{(0)}h_{(2)ij}=&-64\pi G\frac{M-1}{M^2}\psi_{(2-\frac{4}{M})}\phi_{(0)}\notag\\
    \propto&\ (\pi G)^{\frac{M}{2}}\phi^{M}_{(0)}.\label{N=1 metric conformal anomaly}
\end{align}
The order $\rho^0$ of Einstein's equations gives
\begin{align}
    h_{(2)ij}=&-32\pi G\frac{M-1}{M^2}\psi_{(2-\frac{4}{M})}\phi_{(0)}g_{(0)ij},\notag\\
    g^{ij}_{(0)}g_{(2)ij}=&\frac{1}{2}R_{(0)}+\frac{1}{2}g^{ij}_{(0)}F_{(2)ij},\notag\\
    \nabla_{(0)}^{k}g_{(2)ki}=&\partial_{i}(g^{kl}_{(0)}g_{(2)kl})+16\pi G[\frac{2-2M}{M^2}\phi_{(0)}\partial_i\psi_{(2-\frac{4}{M})}+\frac{M^2-2M+2}{M^2}\psi_{(2-\frac{4}{M})}\partial_i\phi_{(0)}]\notag\\
    &+H_{(2)i}+16\pi G[\frac{1}{M}\phi_{(0)}\partial_i\phi_{(2-\frac{4}{M})}+\frac{M-1}{M}\phi_{(2-\frac{4}{M})}\partial_i\phi_{(0)}],
\end{align}
where
\begin{align}
    H_{(2)i}=&\sum_{\substack{m_1+m_2=M\\m_i\in\lbrace{1,2,\cdots,M-1}\rbrace}}\frac{m_2}{M}\Big[\partial_i\Big(g^{kl}_{(\frac{2m_1}{M})}g_{(\frac{2m_2}{M})kl}\Big)-\nabla^k_{(\frac{2m_1}{M})}g_{(\frac{2m_2}{M})ki}\Big]\notag\\
    &+16\pi G\sum_{\substack{m_1+m_2=M-2\\m_i\in\lbrace{1,2,\cdots,M-3}\rbrace}}\frac{m_1+1}{M}\phi_{(\frac{2m_1}{M})}\partial_i\phi_{(\frac{2m_2}{M})}\notag\\
    \propto&\ (\pi G)^\frac{M}{2}\phi_{(0)}^{M-1}\partial_i\phi_{(0)}.
\end{align}
Plugging the FG expansions (\ref{N=1 Fefferman-Graham expansions}) into the regularized on-shell action, we obtain
\begin{align}
I_{\text{reg}}=&-\frac{1}{8\pi G}\int_{\partial\mathcal{M}_{\varepsilon}}d^2x\sqrt{g_{(0)}}\Big[\frac{1}{\varepsilon}-\frac{1}{2}(1-\frac{2}{M})\frac{8\pi G\phi_{(0)}^2}{\varepsilon^{1-\frac{2}{M}}}+\sum_{m=3}^{M-1}\frac{(\pi G)^{\frac{m}{2}}D_{(\frac{2m}{M})}\phi_{(0)}^m}{\varepsilon^{1-\frac{m}{M}}}\notag\\
&+\frac{1}{4}[R_{(0)}+(\pi G)^\frac{M}{2}{D}_{(2)}\phi_{(0)}^M]\text{ln}\varepsilon\Big]+O(1).
\end{align}
Using the invert FG expansion, we can rewrite $g_{(0)ij}$ and $\phi_{(0)}$ in terms of $g_{ij}$ and $\phi$, and finally obtain the allowed counterterm (\ref{N=1 counterterm}) for $N=1$.

\bibliographystyle{JHEP}
\bibliography{reference.bib}

\end{document}